\documentclass[%
onecolumn,
showkeys,
pof,
superscriptaddress
]{revtex4}
\usepackage{bm}
\usepackage{graphicx}
\usepackage{amsmath}
\usepackage{cases}
\usepackage{epstopdf}
\usepackage[caption=false]{subfig}

\begin{document}

\title{Discrete unified gas kinetic scheme on unstructured meshes}

\author{Lianhua Zhu}
\email{lhzhu@hust.edu.cn}
\affiliation{State Key Laboratory of Coal Combustion, Huazhong University of Science and Technology, Wuhan 430074, China}
\author{Zhaoli Guo}
\email{zlguo@hust.edu.cn}
\affiliation{State Key Laboratory of Coal Combustion, Huazhong University of Science and Technology, Wuhan 430074, China}
\author{Kun Xu}
\email{makxu@ust.hk}
\affiliation{Department of Mathematics, Hong Kong University of Science and Technology, Clear Water Bay, Hong Kong, China}

\begin{abstract}
The recently proposed discrete unified gas kinetic scheme (DUGKS) is a 
finite volume method for deterministic solution of the Boltzmann model equation with 
asymptotic preserving property.
In DUGKS, the numerical flux of the distribution function 
is determined from a local numerical solution of the Boltzmann model equation using an unsplitting approach. 
The time step and mesh resolution are not restricted by the molecular collision time and mean free path.
To demonstrate the capacity of DUGKS in practical problems, 
this paper extends the DUGKS to arbitrary unstructured meshes.
Several tests of both internal and external flows are performed, 
which include the cavity flow ranging from continuum to free molecular regimes, a multiscale flow between two connected cavities with a pressure ratio of $10^4$, and a high speed flow over a cylinder in slip and transitional regimes.
The numerical results demonstrate the effectiveness of the DUGKS in simulating multiscale flow problems.

\end{abstract}


\maketitle

\section{Introduction}
Gas flows can be classified into different regimes based on the Knudsen number (Kn), 
which is defined as the ratio of the mean free path of the gas to the physical characteristic length. 
For flows with $\text{Kn} > 0.001$, non-equilibrium effects become important and the classical Navier-Stokes-Fourier (NSF) equations fail to describe such effects \cite{bird94}. 
The Boltzmann equation can serve as a fundamental equation that is valid for the whole range of Knudsen number.

There are mainly two types of numerical approaches for solving the Boltzmann equation. 
The first one is the widely used direct simulation Monte Carlo (DSMC) method \cite{bird94}, which is the prevailing technique for simulation of high-speed rarefied gas flows. However, in DSMC the particle transport and collision processes are decoupled, and the cell size and time step are required to be smaller than the mean free path and the particle collision time, respectively. For flows in near continuum or continuum flow regime, this requirement will lead to enormous computational costs. Another undesired feature of the DSMC is the statistical noise that must be reduced with time consuming sampling and averaging, which will lead to even larger computational costs for low speed flows and transient problems \cite{bird94}. It is noted that some efforts have been devoted to reduce statistical noise of the DSMC methods \cite{fanj01, homolle2007low}.
The second  approach is to solve the Boltzmann equation directly using deterministic numerical schemes. The most popular one of this type is the Discrete Velocity Method (DVM) or Discrete Ordinate Method (DOM) \cite{yangjy95, mieussens00,lizhh09},
 in which the velocity space is discretized into a finite set of discrete velocities and the same  operator splitting technique as DSMC is employed to solve the discrete-velocity kinetic equation \cite{yangjy95,mieussens00,toro2009}.
Therefore, these methods face the same constraints on time-step and cell-size  as the DSMC for the continuum and near-continuum flows. Recently, some asymptotic preserving (AP) schemes have been proposed in order to overcome these disadvantages (\emph{e.g.,} \cite{,bennoune08,filbet10}). These schemes have been shown to be able to recover the Euler solutions in the continuum limit, but it is still not clear whether the Navier-Stokes solutions can be obtained.

Recently, a unified gas kinetic scheme (UGKS) in finite-volume formulation was constructed for all Knudsen number flows \cite{xuk10, xuk11, huangjc12, huangjc13}. 
Unlike the traditional DOM or DVM, the particle transport and collision are considered simultaneously in UGKS in the update of the discrete distribution function. Consequently,  the restriction on the cell size and time step is avoided. Therefore, UGKS can be used to simulate entire Knudsen number flows efficiently \cite{chensz13}. 

A novel discrete unified gas kinetic scheme (DUGKS) for multi-regime flows was proposed recently \cite{guozl13, guozl14}, which shares the same modeling mechanism as the original UGKS \cite{guozl13}.  The main difference lies in the reconstruction of the discrete distribution function at cell-interface. In UGKS, the time-dependent interface distribution function is determined from the local integral solution of the kinetic equation, while in DUGKS, the distribution function at the half time step is determined from a characteristic solution of kinetic equation. This reconstruction includes the coupled effects of particle transport and collision, and makes the updating rule much simplified in comparison with the UGKS.

In previous works \cite{guozl13, guozl14}, the DUGKS has been applied to both low speed and high speed non-equilibrium flows based on structured meshes.
To further demonstrate the potential applications of the DUGKS for practical problems involving complex geometrics or large flow gradients, in this work we will extend the DUGKS to arbitrary unstructured meshes.

The rest of the paper is organized as following. In Sec. 2 the general procedure of the DUGKS on unstructured meshes is presented. In Sec. 3, several numerical examples, including the micro cavity flow, an expansion flow between two connected cavities, and the rarefied gas flow past a circular cylinder, are provided to demonstrate the applicability of the current method in simulating flows at or covering different regimes. A brief summary is given in the last section.

\section{Discrete unified gas kinetic scheme}
\subsection{Shakhov model}
The DUGKS is based on the Boltzmann model equation in which
the collision operator is approximated by the Shakhov model \cite{shakhov68} for monatomic gases. In $D$ dimensional space, the model equation is
\begin{equation}
\frac{\partial f}{\partial t} + \bm \xi \cdot \nabla f = - \frac{1}{\tau}\left[ f - f^S \right],
\label{eq:shak}
\end{equation}
where $f = f(t, \bm \xi, \bm \eta, \bm x)$ is the velocity distribution function of particles 
with velocity $\bm \xi = (\xi_1, \ldots, \xi_D)$ in $D$ dimensional velocity space at position $\bm x = (x_1, \ldots, x_D)$ and time $t$.
Here $\bm \eta = (\xi_{D+1}, \ldots, \xi_3)$ is a vector in a space with dimensionality $L=3-D$, consisting of the rest components of the particle velocity 
$(\xi_1, \xi_2, \xi_3)$ in 3-dimensional space;
$f^S$ is the Shakhov equilibrium distribution function given by the Maxwellian distribution function $f^{eq}$, plus a heat flux correction term as
\begin{equation}
f^S = f^{eq} \left[ 1 + (1-\text{Pr})\frac{\bm c \cdot \bm q}{5pRT}\left( \frac{c^2 + \eta^2}{RT} -5) \right) \right ]
= f^{eq} + f_{Pr},
\label{eq:feq_shak}
\end{equation}
where $\text{Pr}$ is the Prandtl number and $\bm c = \bm \xi - \bm U$ is the peculiar velocity with $\bm U$ being the macroscopic fluid velocity; 
$\bm q$ is the heat flux vector, $R$ is the specific gas constant, and $T$ is the temperature. 
The collision time $\tau$ in Eq.~\eqref{eq:shak} is related to the dynamic viscosity $\mu$ and pressure $p$ by $\tau = \mu/p$. 
The Maxwellian distribution function $f^{eq}$ is given by
\begin{equation}
f^{eq} = \frac{\rho}{(2\pi RT)^{3/2}}\exp\left( - \frac{c^2 + \eta^2}{2RT} \right),
\label{eq:eq_max}
\end{equation}
where $\rho$ is the macroscopic gas density. 
The conservative macroscopic flow variables $\bm W \equiv (\rho, \rho \bm U, \rho E)^T$ are calculated as velocity moments of the distribution function, 
\begin{equation}
\bm W = \int \bm \psi f \text{d}\bm\xi\text{d}\bm\eta ,
\label{eq:mom}
\end{equation}
where $\bm \psi = \left( 1, \bm \xi, \frac{1}{2}(\xi^2 +  \eta^2) \right )^T$ and
$\rho E = \frac{1}{2}(c^2 + \eta^2 ) + C_\text{V}T$ is the total energy with $C_\text{V}$ being the heat capacity at constant volume. 
The heat flux $\bm q$ is defined by 
\begin{equation}
\bm q = \frac{1}{2}\int \bm c (c^2 + \eta^2 )f\text{d}\bm\xi\text{d}\bm\eta.
\label{eq:q}
\end{equation}

The distribution function $f$ depends only on $\bm \xi$ in $D$ dimensional velocity space and is irrelevant to $\bm \eta$. 
To remove the dependence on $\bm \eta$ , two reduced distribution functions can be introduced \cite{yangjy95}
\begin{subequations}
\begin{align}
g(\bm x, \bm \xi , t) = &\int f(\bm \xi, \bm\eta, \bm x, t) \text{d}\bm\eta, 
\label{eq:g} \\
h(\bm x, \bm \xi , t) = &\int \eta^2  f(\bm \xi,\bm\eta, \bm x, t) \text{d}\bm\eta.
\label{eq:h}
\end{align}
\end{subequations}
The macroscopic variables can be computed from these reduced distribution
function as
\begin{equation}
\rho = \int g \text{d}\bm \xi,
\quad \rho \bm U = \int \bm \xi g \text{d}\bm\xi, 
\quad \rho E = \frac{1}{2}\int (\xi^2 g + h) \text{d}\bm \xi,
\label{eq:macro_g}
\end{equation}
and the heat flux can be computed as
\begin{equation}
\bm q = \frac{1}{2}\int \bm c(c^2 g + h) \text{d} \bm\xi.
\label{eq:q_g}
\end{equation}
The evolution equations for the reduced distribution functions can be deduced from Eq.~\eqref{eq:shak} as
\begin{subequations}
\begin{align}
\frac{\partial g}{\partial t} + \bm \xi \cdot \nabla g = &\Omega_h = - \frac{1}{\tau}\left[ g - g^S \right],
\label{eq:g_evo} \\
\frac{\partial h}{\partial t} + \bm \xi \cdot \nabla h = &\Omega_g = - \frac{1}{\tau}\left[ h - h^S \right],
\label{eq:h_evo}
\end{align}
\end{subequations}
where the reduced equilibrium distribution functions $g^S$ and $h^S$ can be deduced from the original equilibrium distribution as
\begin{subequations}
\begin{align}
g^S(\bm x, \bm \xi, t) = & \int f^S(\bm \xi,\bm\eta, \bm x, t) \text{d}\bm\eta = g^{eq} + g_\text{Pr},
\label{eq:gs_evo} \\
h^S(\bm x, \bm \xi, t) = & \int \eta^2 f^S(\bm \xi, \bm\eta, \bm x, t) \text{d}\bm\eta = h^{eq} + h_\text{Pr},
\label{eq:hs_evo}
\end{align}
\label{eq:gh_evo}
\end{subequations}
with 
\begin{subequations}
\begin{align}
g^{eq} = &\frac{\rho}{ (2\pi RT)^{D/2}}\exp\left[ -\frac{c^2}{2RT} \right],
\label{eq:g_eq} \\
h^{eq} = &(3-D)RTg^{eq},
\label{eq:h_eq} \\
g_\text{Pr} = & (1 - \text{Pr}) \frac{\bm c \cdot \bm q}{5pRT}
\left[ \frac{c^2}{RT}-D-2\right]g^{eq},
\label{eq:g_pr}\\
h_\text{Pr} = & (1 - \text{Pr}) \frac{\bm c \cdot \bm q}{5pRT}
\left[\left(\frac{c^2}{RT} -D \right)(3-D)\right]RTg^{eq}.
\label{eq:h_pr}
\end{align}
\end{subequations}

\subsection{Discrete unified gas kinetic scheme on unstructured meshes}
\subsubsection{Updating of the cell-averaged distribution function} 
The updating rules for $g$ and $h$ in Eq.~\eqref{eq:gh_evo} have the same structure,
\begin{equation}
\frac{\partial \phi}{\partial t} + \bm \xi \cdot \nabla \phi = \Omega = - \frac{1}{\tau}\left[ \phi - \phi^S \right].
\label{eq:phi_evo}
\end{equation}
where $\phi=g$ or $h$. The generic symbol $\phi$ will be used to denote $g$ and $h$ in the following.
The DUGKS is an explicit finite volume scheme for solving the kinetic equation \eqref{eq:phi_evo}. 
The computation domain is firstly divided into some control volumes (cells).
By integrating Eq.~\eqref{eq:phi_evo} in each cell from time $t_n$ to $t_{n+1}$,
we have 
\begin{equation}
\phi^{n+1}_j (\bm \xi) - \phi^n_j(\bm \xi) + \frac{\Delta t}{|V_j|}\mathcal{F}_j^{n+1/2}(\bm \xi) 
= \frac{\Delta t}{2} \left [ \Omega_j^{n+1} - \Omega_j^n \right].
\label{eq:phi_evo_d}
\end{equation}
Here $\phi_j$ and $\Omega_j$ is the cell averaged $\phi$ and $\Omega$ in cell $j$; 
$|V_j|$ is the cell volume and $\Delta t = t_{n+1} -t_n$
is the time step. 
Note that the trapezoidal and middle-point rules are used for the collision and convection terms in Eq.~\eqref{eq:phi_evo_d}, respectively.
The term $\mathcal{F}_j^{n+1/2}$ in Eq.~\eqref{eq:phi_evo_d} is the flux of $\phi$ across the interface of cell $j$ and is evaluated as
\begin{equation}
\mathcal{F}_j^{n+1/2}(\bm \xi) = \sum_k \bm \xi \cdot \bm S_j^k \phi_j(\bm x_j^k, \bm \xi, t_{n+1/2}),
\label{eq:flux}
\end{equation}
where $\bm S_j^k$ is the outward normal vector of the $k$th face of cell $j$ with face area $|\bm S_j^k$, and $\bm x_j^k$ is the center of the face.
Equation\eqref{eq:phi_evo_d} can be rewritten in an explicit form by introducing the transformed distribution functions \cite{guozl13, guozl14},
$\tilde{\phi}$ and $\tilde{\phi}^+$ 
\begin{equation}
\tilde{\phi}_j^{n+1} = \tilde{\phi}_j^{+,n} + \mathcal{F}_j^{n+1/2},
\label{eq:phi_tilde_evo}
\end{equation}
where 
\begin{subequations}
\begin{align}
\tilde{\phi}& = \phi - \frac{\Delta t}{2}\Omega = \frac{2\tau + \Delta t}{2\tau}\phi - \frac{\Delta t}{2\tau}\phi^S
\label{eq:phi_tilde}, \\
\tilde{\phi}^+& = \phi + \frac{\Delta t}{2}\Omega = \frac{2\tau - \Delta t}{2\tau + \Delta t}\tilde{\phi} + \frac{2 \Delta t}{2\tau + \Delta t}\phi^S.
\label{eq:phi_tilde_plus}
\end{align}
\end{subequations}
Due to the conservative property of the collision term, the conservative variables can also be calculated from
the transformed distribution functions $\tilde{\phi}$  as \cite{guozl14}
\begin{equation}
\rho = \int \tilde{g} \text{d}\bm \xi,
\quad \rho \bm U = \int \bm \xi \tilde{g} \text{d} \bm\xi, 
\quad \rho E = \frac{1}{2}\int (\xi^2 \tilde{g} + \tilde{h}) \text{d} \bm \xi,
\label{eq:macro_tg}
\end{equation}
and 
\begin{equation}
\bm q = \frac{2\tau}{2\tau+\Delta t\text{Pr}} \tilde{\bm q}, ~ \text{with} ~ 
\tilde{\bm q} = \frac{1}{2}\int\bm c(c^2 \tilde g + \tilde h) \text{d}\bm\xi.
\label{q_form_tilde}
\end{equation}
Therefore, in actual implementation, the evolution of transformed distribution functions $\tilde{\phi}$ is tracked according to Eq.~\eqref{eq:phi_tilde_evo},
instead of the original distribution functions $\phi$ in order to avoid implicit computations. This is one of the major differences between the DUGKS and the UGKS methods. 

\subsubsection{Flux evaluation on unstructured mesh}
\begin{figure}[htbp]
\centering
\includegraphics[width=0.3\textwidth]{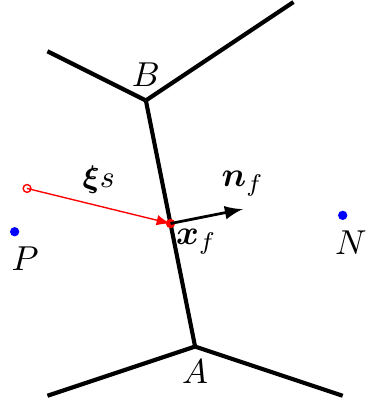}
\caption{Flux evaluation}
\label{fig:face}
\end{figure}

To update $\tilde{\phi}_j$ according to Eq.~\eqref{eq:phi_tilde_evo}, 
the flux $\mathcal F_j^{n+1/2}$ is required.
From the definition of $\mathcal F_j^{n+1/2}$ given by Eq.~\eqref{eq:flux},
the original distribution functions at middle time step at cell interfaces, 
\emph{i.e.}, $\phi^{n+1/2}(\bm x_j^k, \bm \xi)$ is needed. 
This is done by solving the kinetic equation\eqref{eq:phi_evo} locally around the cell interface. To this end, Eq.~\eqref{eq:phi_evo} is integrated from time $t_n$ to $t_{n+1/2}$ along a characteristic line which ends at the face center $\bm x_f$,
\begin{equation}
\phi ^{n+1/2}(\bm x_f, \bm \xi)- \phi^n( \bm x_f - \bm \xi s, \bm \xi) 
= \frac{s}{2}\left[ \Omega^{n+1/2}(\bm x_f, \bm \xi) + \Omega^n (\bm x_f-\bm \xi s, \bm \xi) \right],
\label{eq:phi_half_evo}
\end{equation}
where $s=t_{n+1/2} - t_n$ is the half time step. Here the trapezoidal rule is used again for the collision term.
Similar to the treatment of Eq.~\eqref{eq:phi_evo_d}, another two transformed distribution functions are introduced as
\begin{subequations}
\begin{align}
\bar{\phi} & = \phi - \frac{s}{2}\Omega = \frac{2\tau + s}{2\tau}\phi - \frac{2}{2\tau}\phi^S,
\label{eq:phi_bar}\\
\bar{\phi}^+ & = \phi + \frac{s}{2}\Omega = \frac{2\tau - s}{2\tau + s}\bar{\phi} - \frac{2s}{2\tau + s}\phi^S
\label{eq:phi_bar_plus},
\end{align}
\end{subequations}
then Eq.~\eqref{eq:phi_half_evo} can be expressed explicitly as
\begin{equation}
\bar{\phi}^{n+1/2}(\bm x_f, \bm \xi) = \bar{\phi}^{+,n}(\bm x_f - \bm \xi s,\bm \xi) .
\label{eq:phi_bar_evo_d}
\end{equation}
Piecewise linear reconstruction in the upstream neighboring cells are employed to interpolate $\bar{\phi}^{+,n}(\bm x_f - \bm \xi s)$
from the cell centered $\bar{\phi}^{+,n}$, where the neighboring cells are identified by the direction of the particle velocity $\bm \xi$.
To demonstrate this procedure, here we consider a general case as illustrated in Fig.~\ref{fig:face}.
$AB$ is a cell interface with its center locating at $\bm x_f$ and the unit normal vector $\bm n_f$ pointing from cell $P$ to cell $N$.
The distribution function $\bar{\phi}^+(\bm x_f - \bm \xi s, \bm \xi, t_n)$ is evaluated as
\begin{equation}
\bar{\phi}^+(\bm x_f -\bm \xi s, \bm \xi) = 
\bar{\phi}^+(\bm x_C, \bm \xi) + (\bm x_f - \bm x_C - \bm \xi s) \psi(\bm x_C, \bm \xi) \cdot{\nabla \bar{\phi}^+(\bm x_C, \bm \xi)},
\label{eq:upwind}
\end{equation}
where $C$ stands for $P$ if $\bm \xi \cdot \bm n_f > 0$, or $N$ otherwise. 
The gradient $\nabla\bar{\phi}^+$ at the cell center is determined using the least square method. For instance, the gradient of cell $P$ is evaluated as
\begin{equation}
\left(\nabla\bar{\phi}^+ \right)_P = \sum_i \omega_i^2 \bm G^{-1}\cdot \bm d_i\left[ \left ( \bar{\phi}^+\right)_P - \left(\bar{\phi}^+\right)_{N_i} \right],
\end{equation}
where the tensor $\bm G$ is defined as
\begin{equation}
\bm G = \sum_i \omega_i^2 \bm d_i \bm d_i,
\end{equation}
with $\bm d_i$ being the spatial vector from  $P$ to its $i$th adjacent cell center $N_i$, and $\omega_i = 1/|d_i|$ being the weighting factor. The function $\psi(\bm x_C, \bm \xi)$ in Eq.~\eqref{eq:upwind} denotes the gradient limiter which is used to suppress numerical oscillations in regions with discontinuities, such as the shock layer in continuum regime.
In this work, we adopt the Venkatakrishnan limiter \cite{venkatakrishnan95} which is a typical one for flow computations on unstructured meshes.

The time step in the DUGKS is determined by the Courant-Friedrichs-Lewy (CFL) condition,
\begin{equation}
\Delta t = \alpha \left( \frac{\Delta x}{ |\bm U| + |\bm \xi|} \right)_{\text{min}},
\end{equation}
where $0 < \alpha < 1$ is the CFL number. $\Delta x$ is the distance between the centers of two neighboring cells that share an interface. 

After getting $\bar{\phi}$ at face centers according to Eqs.~\eqref{eq:phi_bar_evo_d} and\eqref{eq:upwind},
the original distribution functions $\phi$ can be recovered from Eq.~\eqref{eq:phi_bar}. 
The macro variables at time $t_{n+1/2}$ that used to evaluate the equilibrium distribution functions $\phi^S$ 
are calculated from $\bar{\phi}$ as
\begin{equation}
\rho = \int \bar{g} d \bm \xi,
\quad \rho \bm U = \int \bm \xi \bar{g} d \bm \xi, 
\quad \rho E = \frac{1}{2}\int (\xi^2 \bar{g} + \bar{h}) d \bm \xi,
\label{eq:macro_gbar}
\end{equation}
and 
\begin{equation}
\bm q = \frac{2\tau}{2\tau+s\text{Pr}} \bar{\bm q}, ~ \text{with} ~ 
\bar{\bm q} = \frac{1}{2}\int\bm c(c^2 \bar g + \bar h)d \bm \xi.
\label{q_form_bar}
\end{equation}
Then the flux across each cell interface can be evaluated according to Eq.~\eqref{eq:flux}. Finally, the cell centered $\tilde{\phi}$ can be advanced to the new time level according to Eq.~\eqref{eq:phi_tilde_evo}.

The updating procedures presented above are all based on continuous velocity space for convenience.
In actual implementation, the continuous velocity space is discretized into a finite discrete velocity set
$\{{\bm \xi_i}\}$ like the DVM \cite{yangjy95}, and the distribution functions such as $\tilde{g}$ and $\tilde{h}$ are defined at these discrete velocity points as $\tilde{g}_i$ and $\tilde{h}_i$.
Proper quadrature rule such as the Newton-Cotes and Gauss-Hermite, is used to approximate the moments,
\begin{equation}
\rho = \sum_i \varpi_i \tilde{g}_i, \quad \rho\bm U = \sum_i \varpi_i \bm \xi_i \tilde{g}_i, \quad
\rho E = \frac{1}{2}\sum_i \varpi_i \left[ \xi_i^2 \tilde{g}_i + \tilde{h}_i \right],
\label{eq:quadrature}
\end{equation} 
where the $\varpi_i$ are the weight coefficients for the corresponding quadrature rule.

\section{Numerical examples}
We will apply the proposed DUGKS on unstructured meshes to two internal and one external flows to demonstrate its performance in multiscale flow simulations.
The first one is the two dimensional lid driven cavity flow at different flow regimes.
The second one is a multiscale unsteady gas expansion problem in which the Knudsen number ranges from $10^{-3}$ to $10$.
The last one is a supersonic rarefied gas flow with Mach number $\text{Ma=5}$ passing through a circular cylinder at $\text{Kn} = 0.1$ and $1$.

As the DUGKS is an explicit scheme, the simulations start from an equilibrium state based on given initial macro fields.
For steady problems, such as the first or the last test case, the flow fields evolve into the final steady states, which is defined as 
the average relative change of the temperature field in two successive steps
being less than $10^{-8}$, \emph{i.e.},
\begin{equation}
\varepsilon^n = \frac{\sum_i |T_i^{n+1} - T_i^n| }{\sum_i T^n_i} < 10^{-8},
\label{eq:steady}
\end{equation}
where the summations are taken over all of the cells.

In all of the tests the simulated gas is argon, with molecular mass $m=6.63^{-26}\text{kg}$ and molecular diameter $d = 4.17\times 10^{-10}\text{m}$. The viscosity of the gas is assumed to depend on temperature following a power-law,
\begin{equation}
\mu=\mu_{ref}\left(\dfrac{T}{T_{ref}}\right)^{\omega}, 
\end{equation}
where $\mu_{ref}$ is the viscosity at the reference temperature $T_{ref}$. 
Here we choose $\omega=0.81$, and the referenced viscosity is set to be that of a  hard-sphere gas, as used in DSMC \cite{bird94}.

\subsection{Cavity flow}
The two dimensional lid driven cavity flow is a standard benchmark problem for the validation of classical CFD methods in continuum regime.
This problem has also been studied recently by Benzi\emph{et~al.} \cite{john11} using a parallel DSMC code at Knudsen numbers $\text{Kn} = 10, 1.0, 0.075$
and was later used as an benchmark test case to validate the UGKS and DUGKS in a wide range of flow regimes \cite{huangjc13,chensz13,guozl13,wangp15}.
To demonstrate that the DUGKS can recover the Navier-Stokes limit without resolving the mean free path scale in the continuum regime, 
we also simulate this case in full range of Knudsen numbers.

The flow domain is a square cavity with length $L=1\text{m}$.
The upper wall moves with a constant velocity $U_\text{w}$, while other walls are kept fixed.
The temperature at the four walls is fixed at $T_\text{w}=273\text{K}$ and is used as the referenced temperature.
The walls are assumed to be fully diffusive and the boundary conditions are realized following the method presented in Refs. \cite{guozl13, guozl14}.
The Knudsen number is defined as $\text{Kn}=\lambda/L$, where $\lambda$ is the mean-free-path of the gas.
Different Knudsen number can be achieved by adjusting the initial density $\rho_{ref}$.

Both rarefied and continuum flows are simulated. In rarefied regimes, three values of the Knudsen number, $\text{Kn}=10, 1$ and $0.075$, are considered.  
The velocity of the upper wall is set to be $U_\text{w}=50\text{m/s}$,
which is the same configuration as used in the DSMC and UGKS simulations \cite{john11, huangjc13}. For continuum flows, two Reynolds numbers are considered, i.e.,  $\text{Re}=400$ and $1000$. Here the Reynolds number is defined as $\mbox{Re}=\rho_{ref} L U_w/\mu_{ref}$, and the corresponding Knudsen numbers are $3.7763\times 10^{-4}$ and $1.5105\times 10^{-4}$, respectively. Furthermore, the Mach number $\text{Ma} = U_W/\sqrt{\gamma RT_{ref}}= 0.1$, so that the flow is nearly incompressible and we can compare our results with the benchmark solutions \cite{ghia82} based on the incompressible Navier-Stokes equations.

\begin{figure}[htbp]
\centering
\subfloat[]{\includegraphics[width=0.35\textwidth]{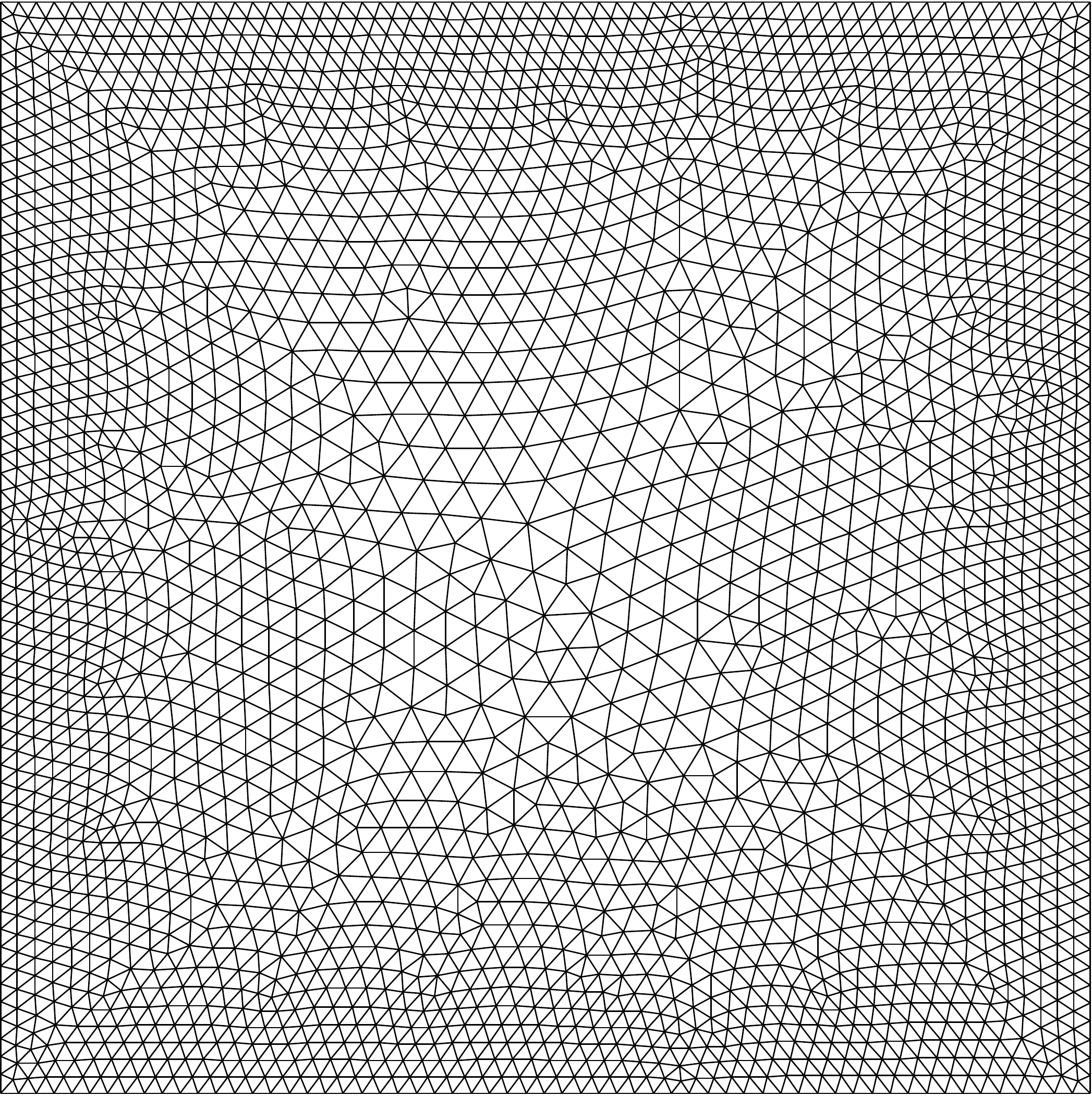}}
\quad
\subfloat[]{\includegraphics[width=0.35\textwidth]{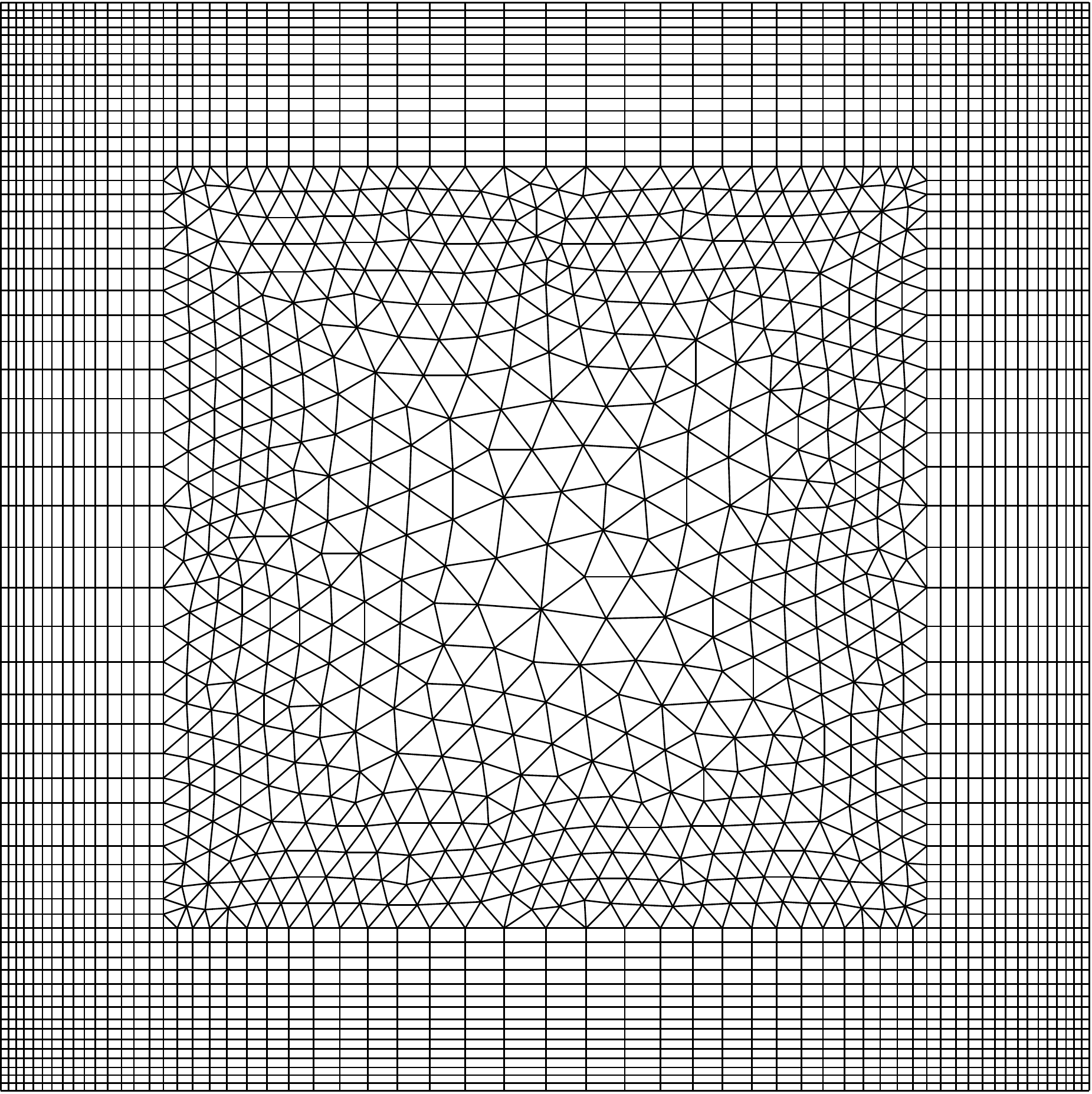}}
\caption{Meshes for the cavity flow.
(a) $\text{Kn}=10,1$ and $0.075$.
(b) $\text{Re}=400$ and $1000$. }\label{fig:cavity_meshes}
\end{figure}

In previous work \cite{guozl13}, the DUGKS with structured meshes has been employed to simulate the cavity flow at different flow regimes. 
Here we choose unstructured meshes to demonstrate the performance of the proposed method.
Figure~\ref{fig:cavity_meshes} presents the meshes used for flows with finite Knudsen numbers and continuum flows, respectively.
Note that the mesh in Fig.~\ref{fig:cavity_meshes}(b) is a hybrid mesh with quadrilateral cells near the walls, 
which performs better than a pure triangular mesh in capturing the boundary layer effect that is important in continuum flows.

The discretization of velocity space and quadrature rules are chosen dependent on the Knudsen number. For highly rarefied flows, (\emph{i.e.}, $\text{Kn}=10, 1$), 
We will use the Newton-Cotes rule with $101\times101$ velocity points distributed uniformly in the range of $[-4\sqrt{2RT_\text{w}},4\sqrt{2RT_\text{w}}] \times[-4\sqrt{2RT_\text{w}},4\sqrt{2RT_\text{w}}]$.
For the case of $\text{Kn}=0.075$, we will adopt the half-range Gauss-Hermit quadrature  with $28\times28$ velocity points.
For continuum flows, we will employ the half-range Gauss-Hermit quadrature rule with $16 \times 16$ velocity points. The CFL number is fixed at 0.8 in all simulations.

\begin{figure}[htbp]
\centering
\subfloat[]{\includegraphics[width=0.315\textwidth]{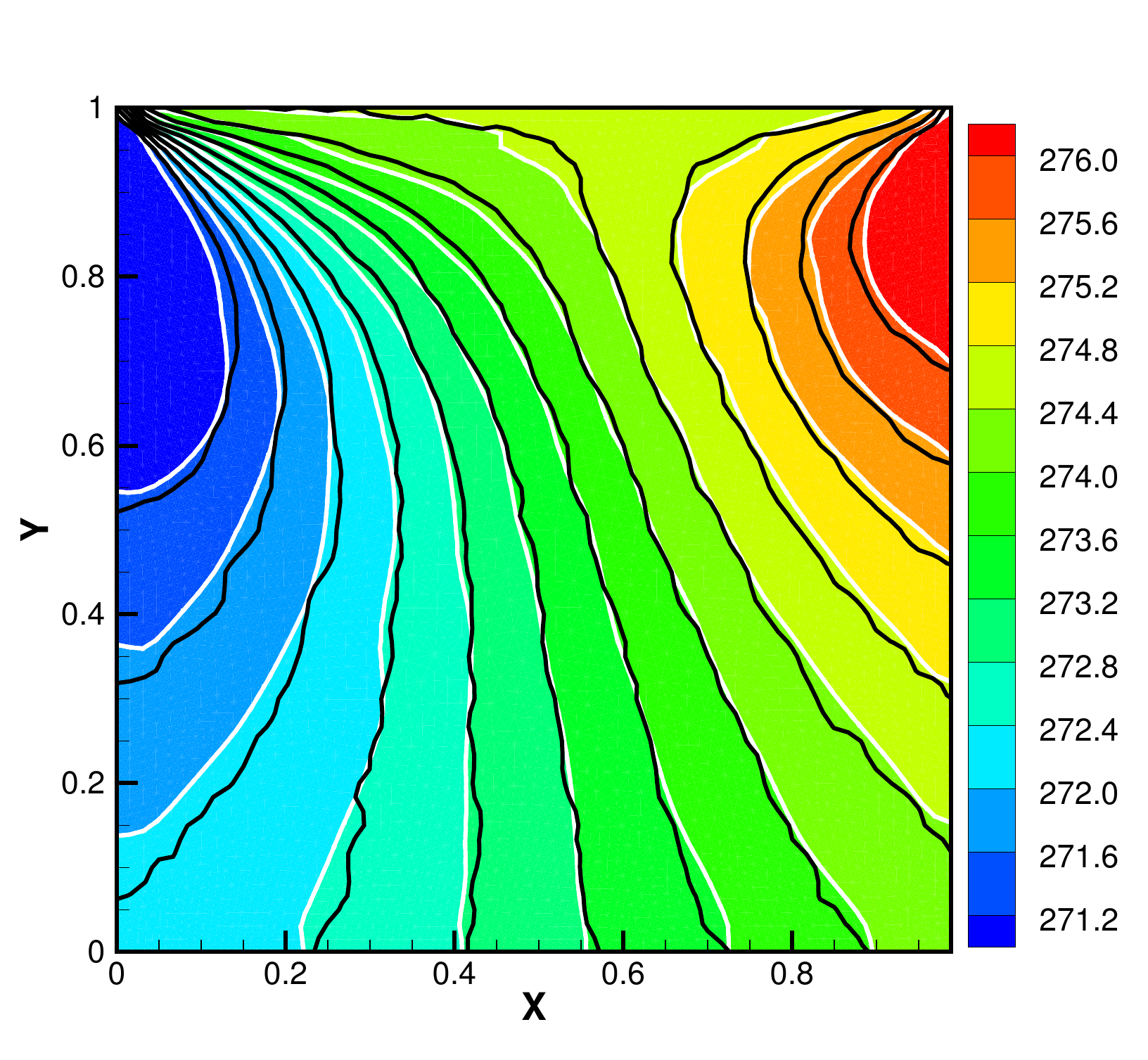}}~
\subfloat[]{\includegraphics[width=0.29\textwidth]{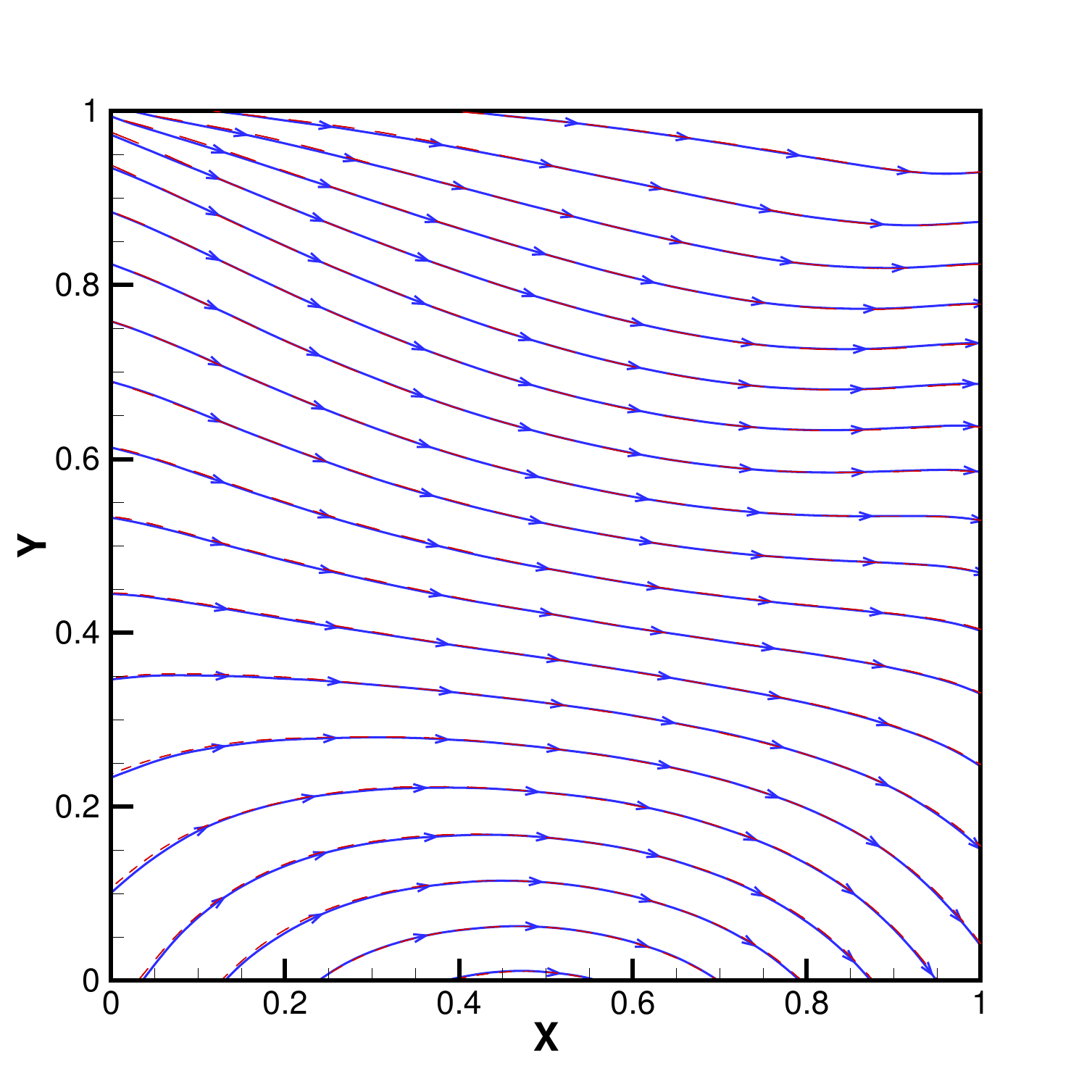}}~
\subfloat[]{\includegraphics[width=0.29\textwidth]{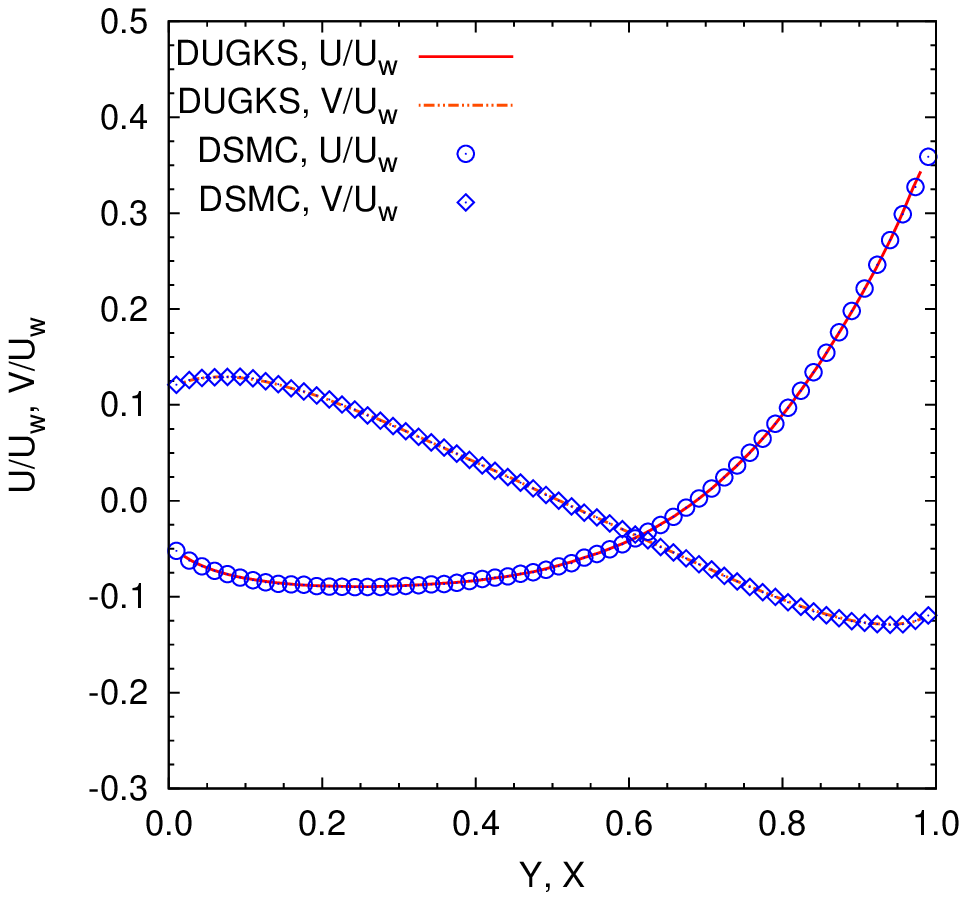}}
\caption{Results of the cavity flow at $\text{Kn}=10$. (a) temperature contours, black line: DSMC, white line with background: DUGKS,
(b) heat flux: blue solid line with arrows: DSMC, red dashed line : DUGKS, (c) U-velocity along vertical center line and 
V-velocity along horizontal central line}\label{fig:Kn10}
\end{figure}

\begin{figure}[htbp]
\centering
\subfloat[]{\includegraphics[width=0.315\textwidth]{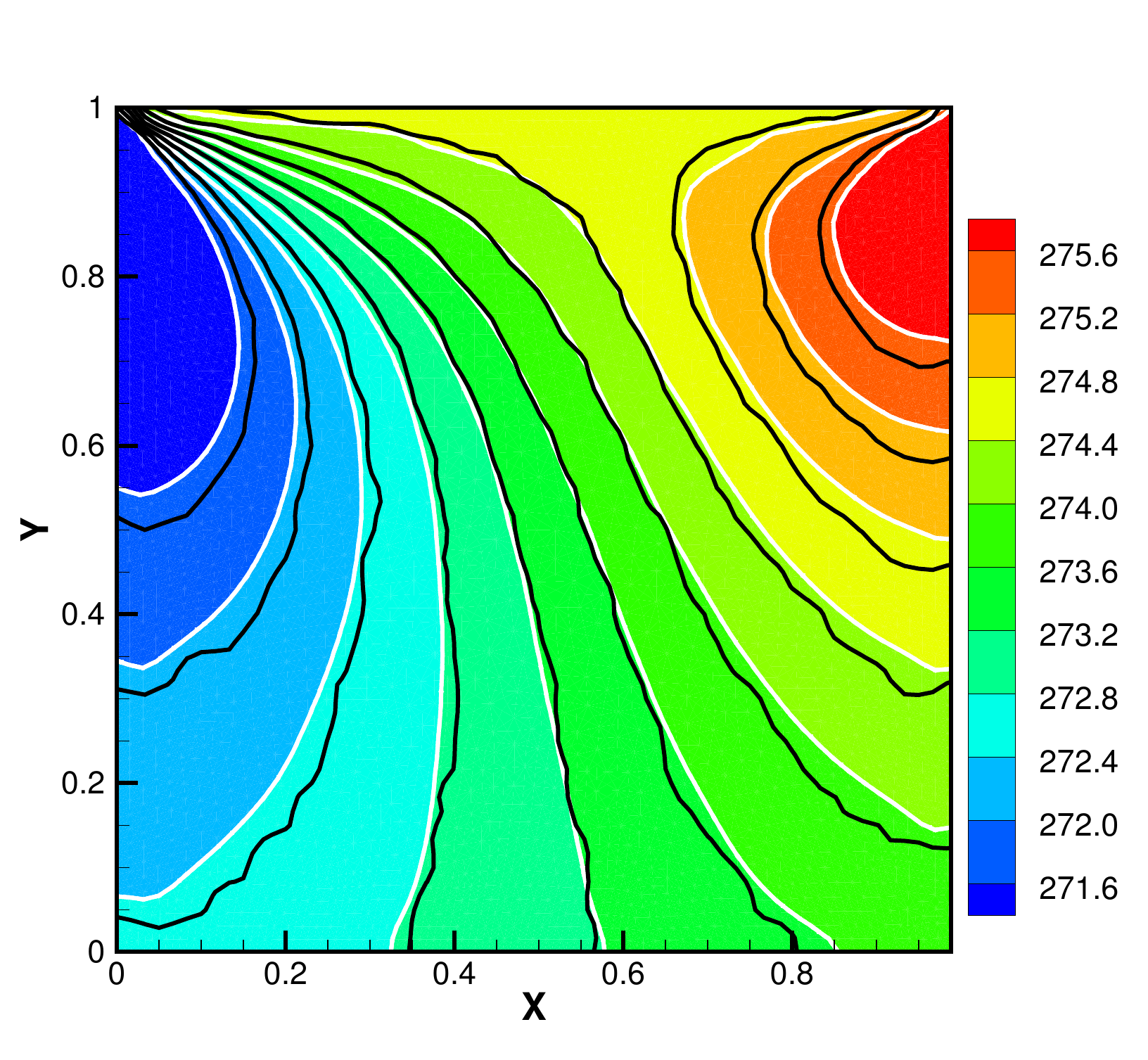}}~
\subfloat[]{\includegraphics[width=0.29\textwidth]{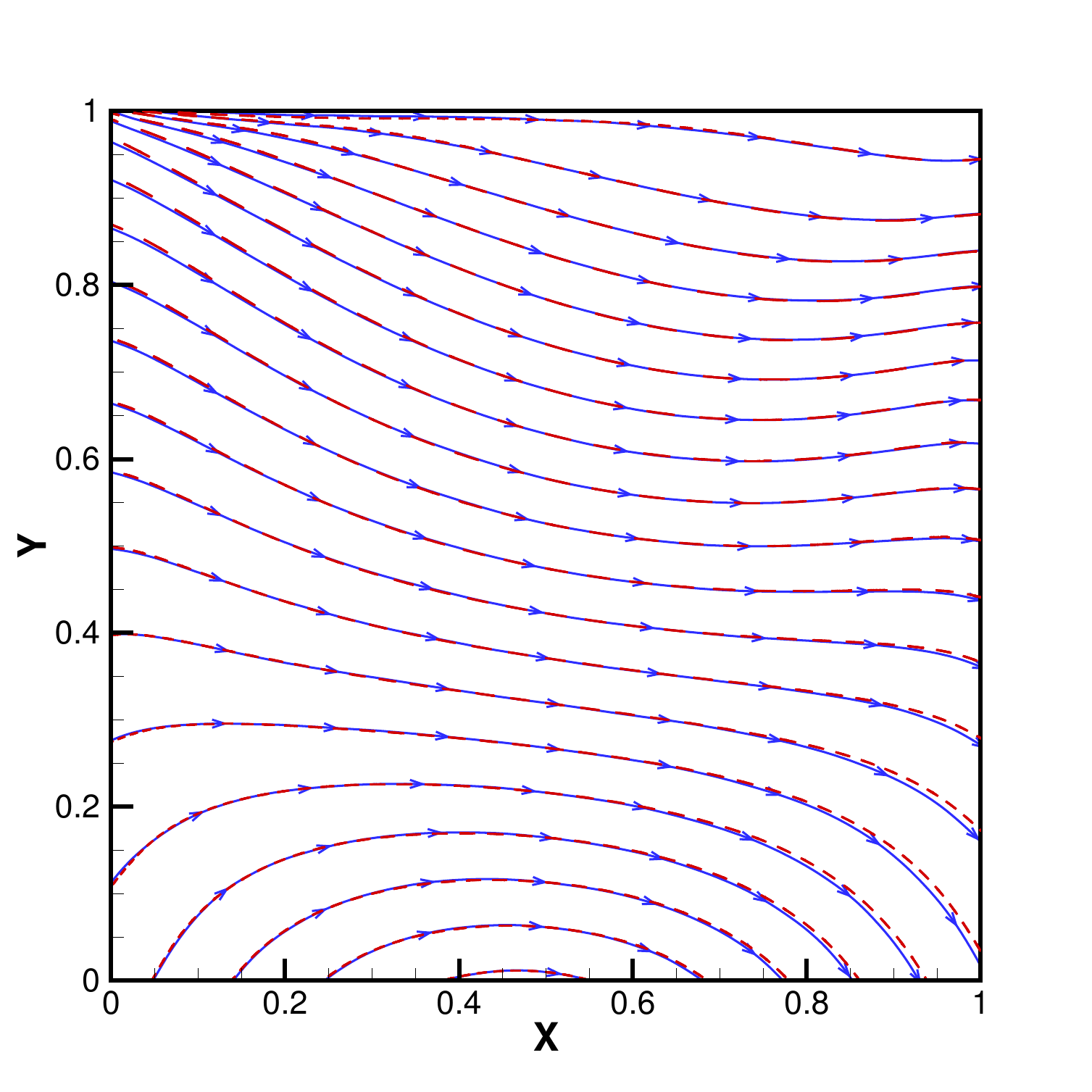}}~
\subfloat[]{\includegraphics[width=0.29\textwidth]{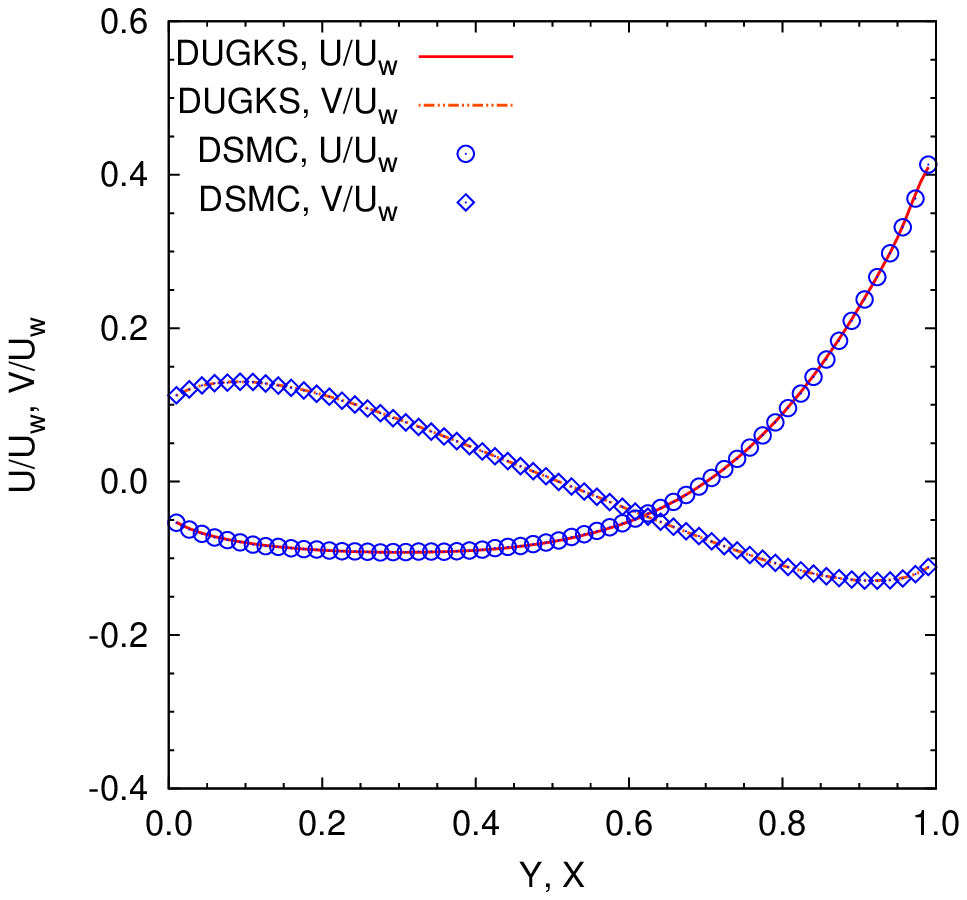}}
\caption{Results of the cavity flow at $\text{Kn}=1$. (a) temperature contours, black line: DSMC, white line with background: DUGKS,
(b) heat flux: blue solid line with arrows: DSMC, red dashed line : DUGKS, (c) U-velocity along vertical center line and 
V-velocity along horizontal central line}\label{fig:Kn1}
\end{figure}

\begin{figure}[htbp]
\centering
\subfloat[]{\includegraphics[width=0.315\textwidth]{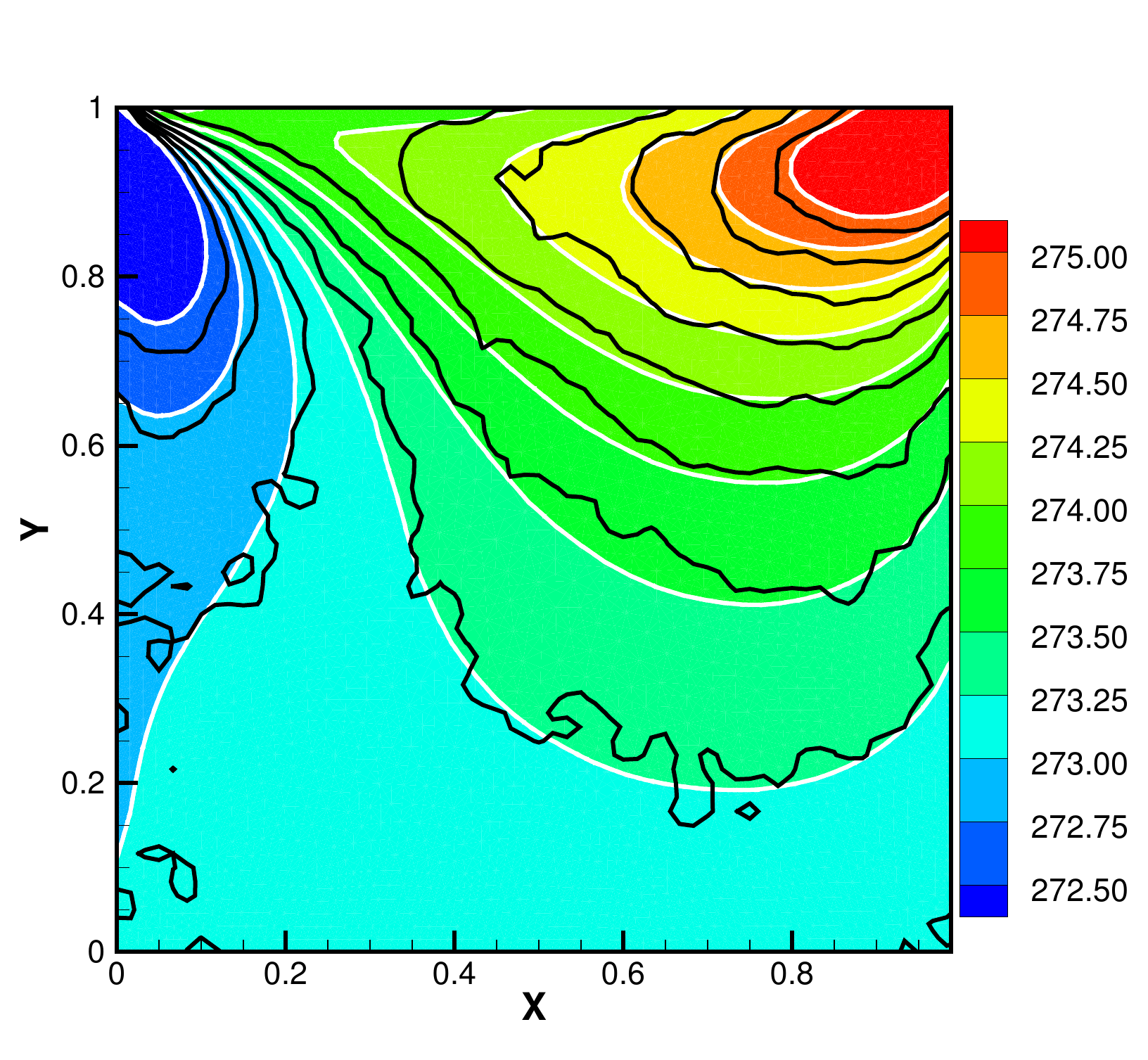}}~
\subfloat[]{\includegraphics[width=0.29\textwidth]{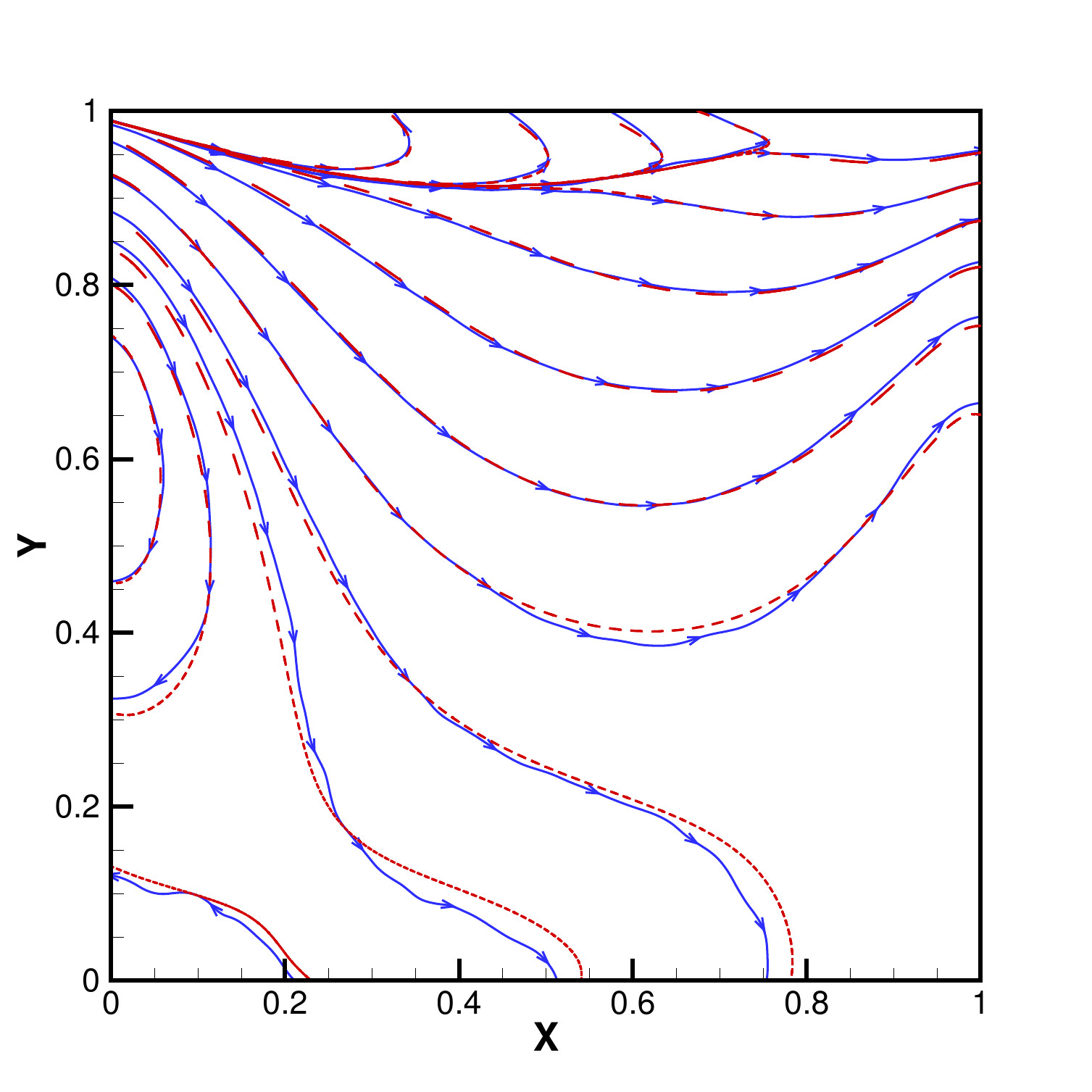}}~
\subfloat[]{\includegraphics[width=0.29\textwidth]{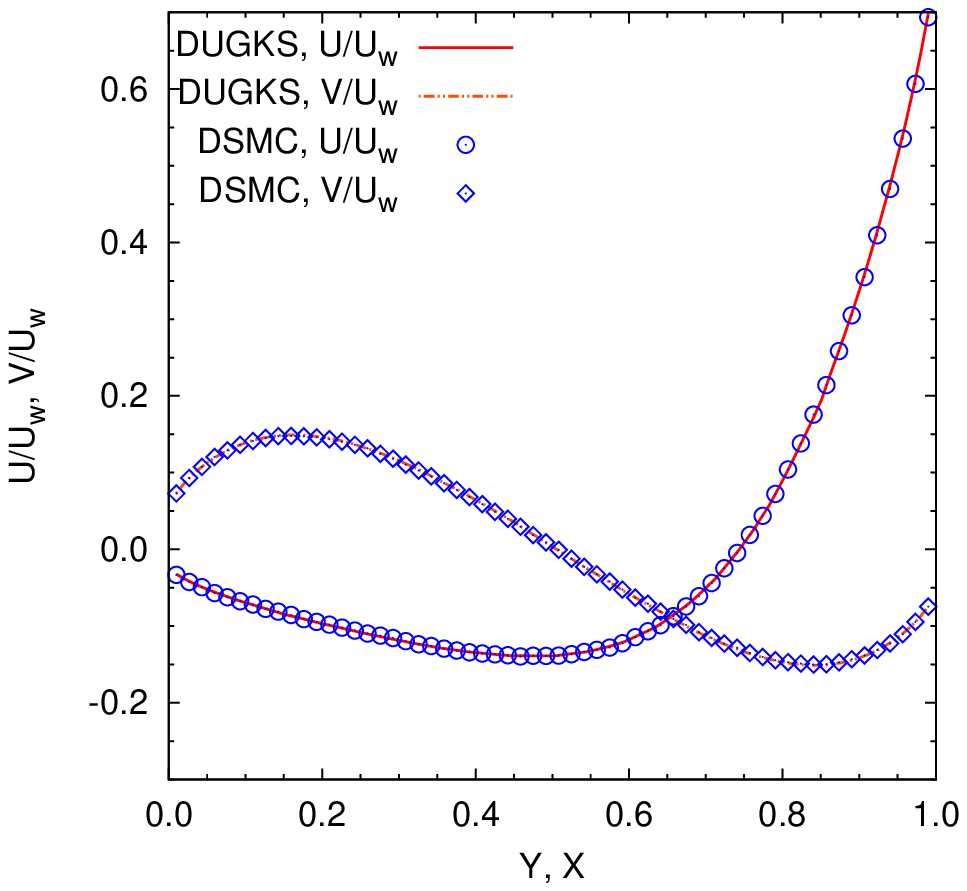}}
\caption{Results of the cavity flow at $\text{Kn}=0.075$. (a) temperature contours, black line: DSMC, white line with background: DUGKS,
(b) heat flux: blue solid line with arrows: DSMC, red dashed line : DUGKS, (c) U-velocity along vertical center line and 
V-velocity along horizontal central line}\label{fig:Kn0.075}
\end{figure}

Figures~\ref{fig:Kn10}-\ref{fig:Kn0.075} present the temperature field, heat flux, and velocity $(U, V)$ on the vertical and horizontal center lines, together with the DSMC solutions for the cases of $\text{Kn}=10,1$ and $0.075$, respectively.
It can be seen that the present results agree well with DSMC results. It is interesting to note that the direction of the heat flux is inconsistent with the temperature gradient in each case, suggesting that the Fourier law breaks down, even at the Knudsen number as small as 0.075. 

\begin{figure}[htbp]
\centering
\subfloat[]{\includegraphics[width=0.44\textwidth]{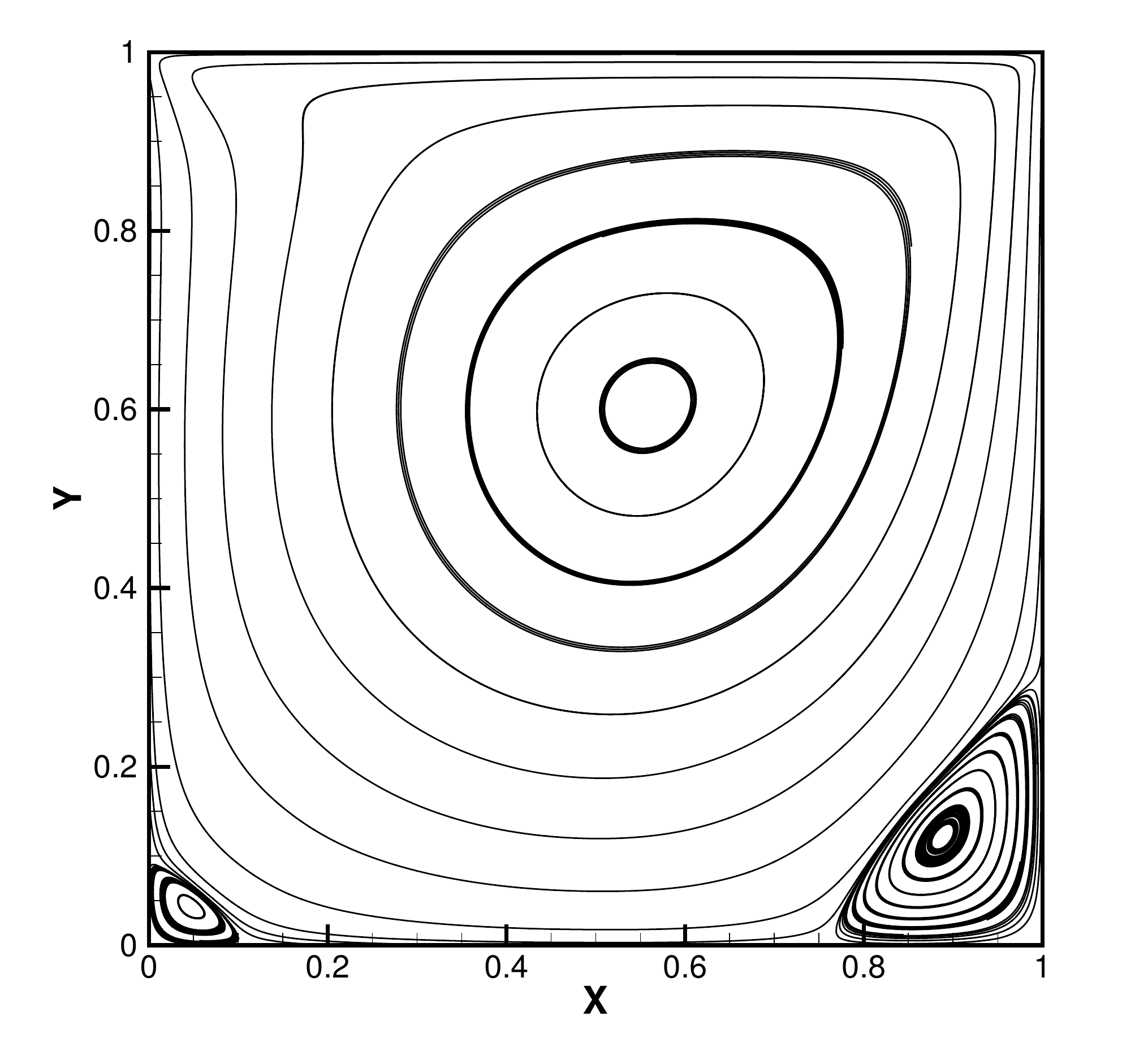}}~
\subfloat[]{\includegraphics[width=0.42\textwidth]{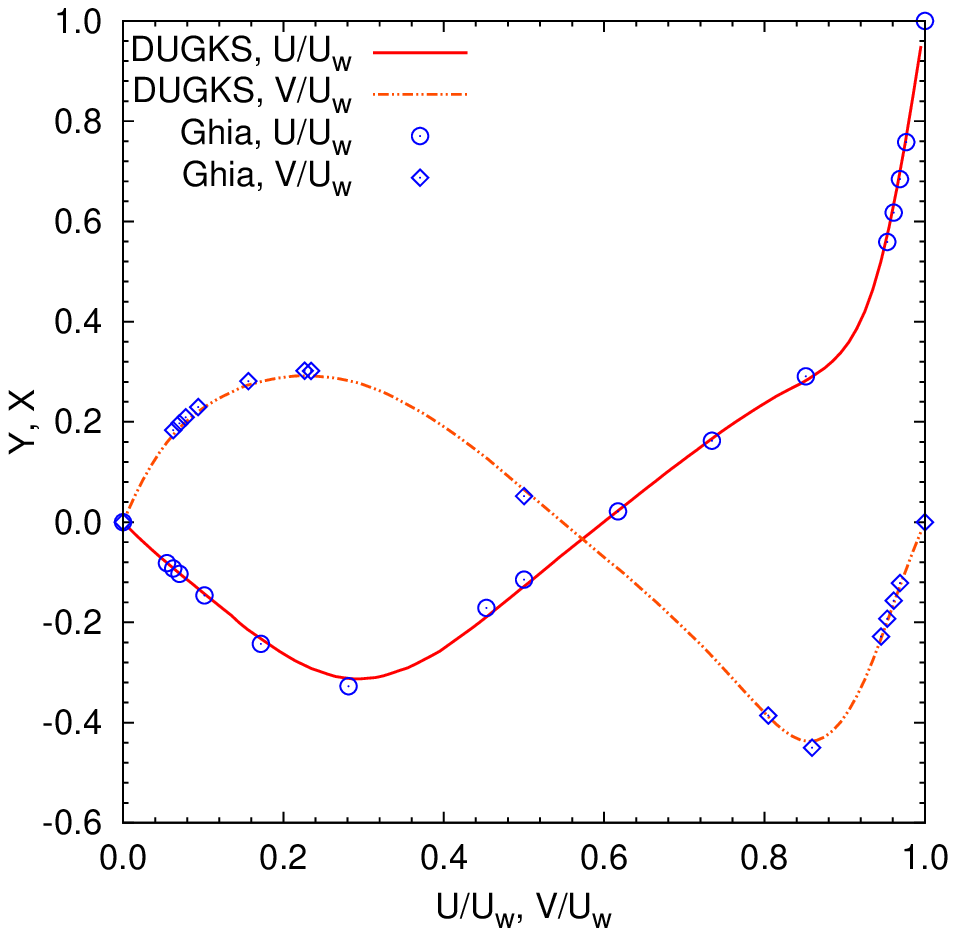}}
\caption{
Results of the cavity flow at $\text{Re}=400$, $\text{Kn}=3.7763\times 10^{-4}$.
(a) Velocity streamline
(b) U-velocity alone vertical central line and V-velocity alone horizontal central line
}\label{fig:Re400}
\end{figure}

\begin{figure}[htbp]
\centering
\subfloat[]{\includegraphics[width=0.44\textwidth]{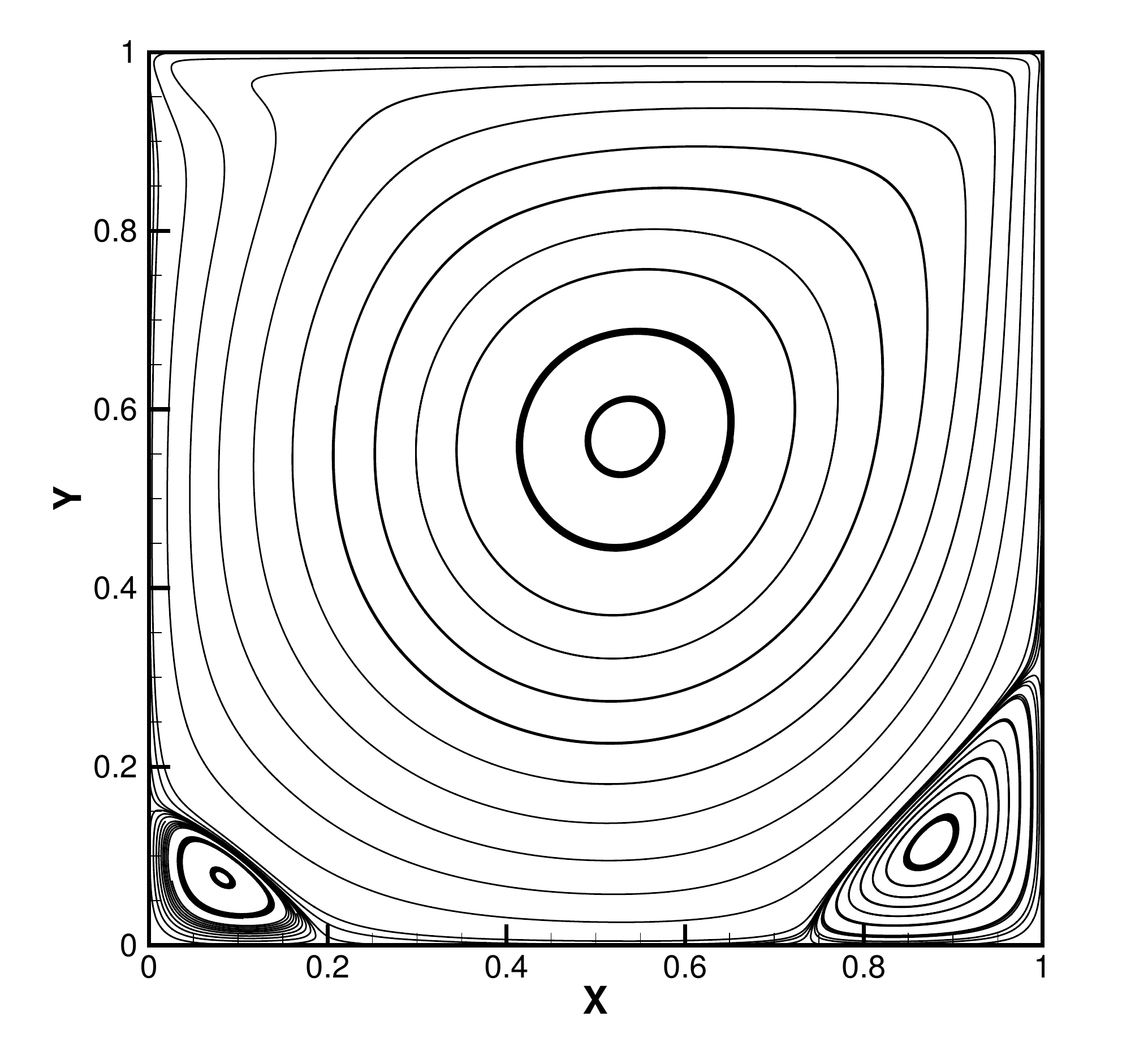}}~
\subfloat[]{\includegraphics[width=0.42\textwidth]{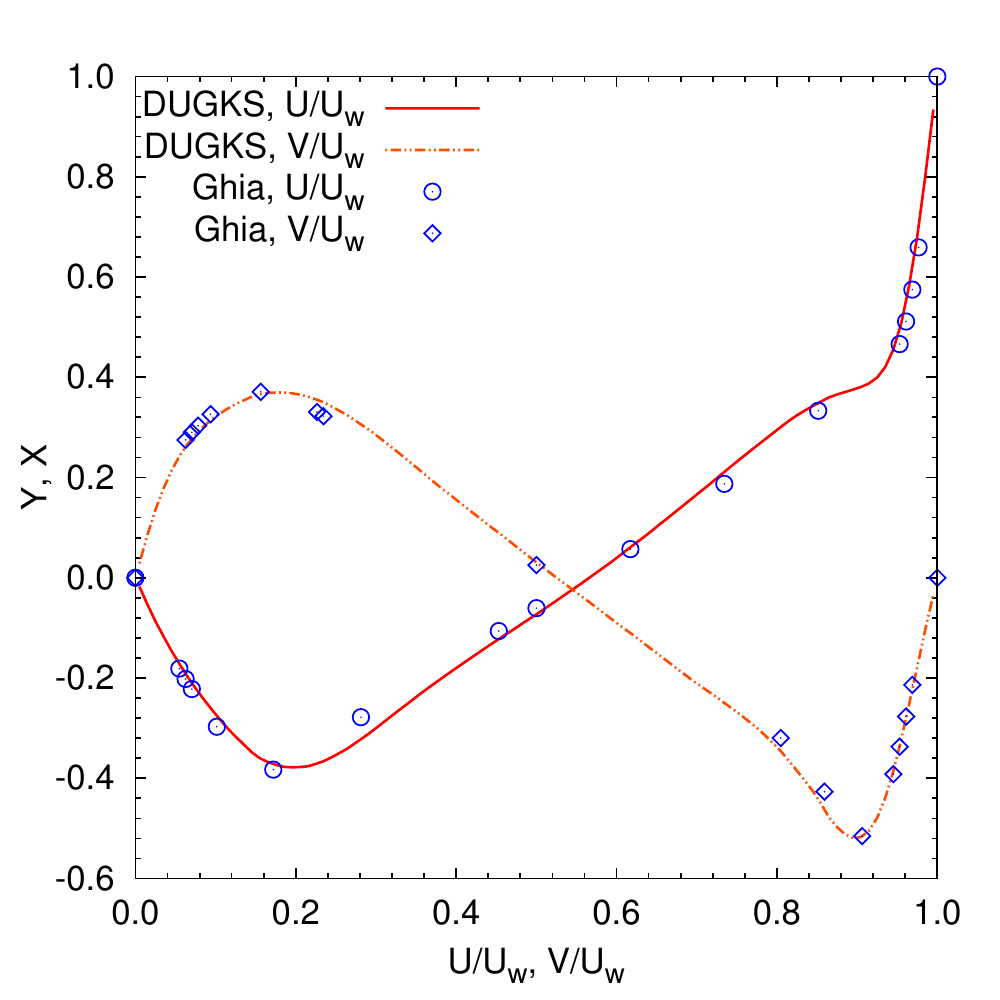}}
\caption{
Results of the cavity flow at $\text{Re}=1000$, $\text{Kn}=1.5105\times 10^{-4}$.
(a) Velocity streamline
(b) U-velocity alone vertical central line and V-velocity alone horizontal central line
}\label{fig:Re1000}
\end{figure}

Figures~\ref{fig:Re400} and \ref{fig:Re1000} show the streamline and velocity profiles for the cases of $\text{Re}=400$ and $1000$ respectively.
The benchmark solutions \cite{ghia82} are also included for comparison. 
We can see that even though the cell size is much larger than the mean free path in these cases,the DUGKS results are still in very close agreement with the benchmark data. So the DUGKS recovers the Navier-Stokes solutions in the continuum limit. We would also like to point out that for most traditional DVM methods, 
the numerical dissipation is proportional to cell size due to the splitting treatment of particle transport and collision processes. 
This may lead to significant errors for unstructured meshes as the cell size changes dramatically. The above results indicate that the DUGKS can avoid this difficulty with the coupled treatment of particle transport and collision.

\subsection{Multiscale expansion flow between two connected cavities}
\begin{figure}[htbp]
\centering
\includegraphics[width=0.8\textwidth]{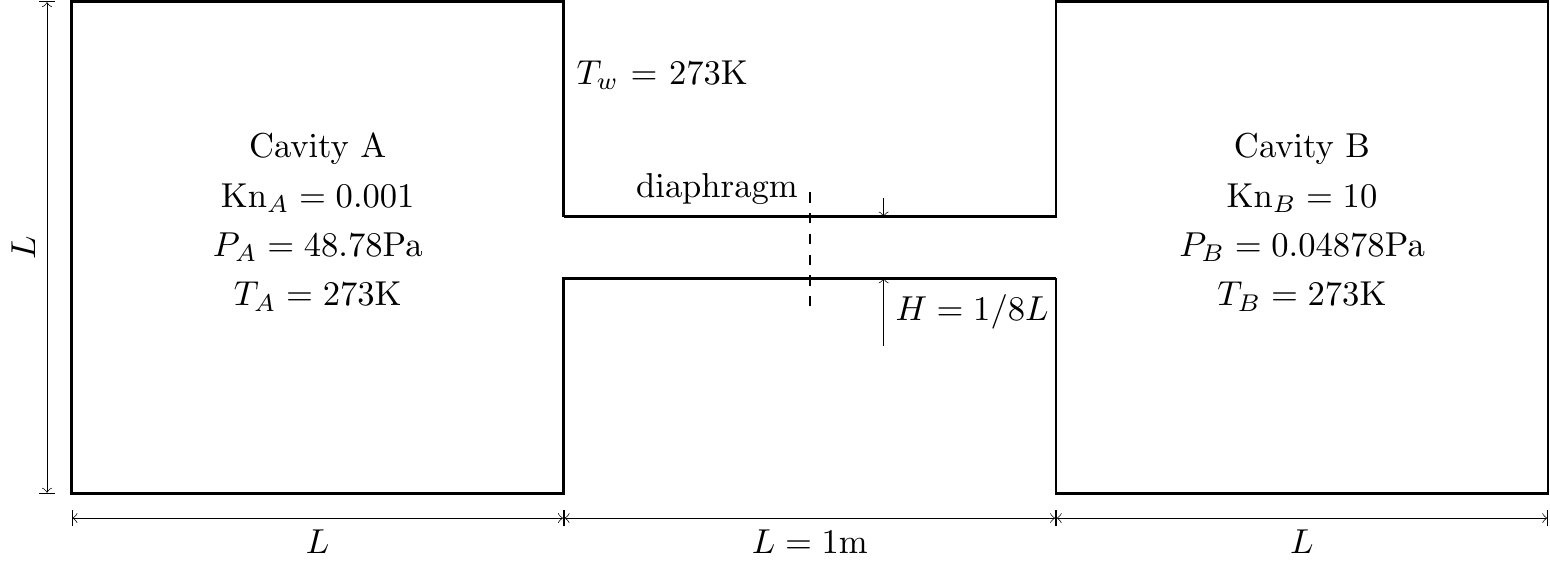}
\caption{Gas expansion between two cavities connected by a channel}
\label{fig:expansion}
\end{figure}

In the above subsection, the cavity flow at specific regimes have been simulated.
Now we consider a gas expansion flow between two connected cavities with different initial pressures. This flow is an unsteady multiscale problem where different flow regimes appear in a single run.
The flow configuration is sketched in Fig.~\ref{fig:expansion}. 
Two square cavities $A$ and $B$ connected by a channel are initially maintained at different pressure and separated by a diaphragm at the middle of the channel.
The height of the cavity is $L=1$m, and the length and width of channel are $L$ and $H$ with $H=L/8$.
Initially, the temperature of the gas in the system and that of the solid walls are maintained at $273$K, which is used as the reference temperature. 
The initial Knudsen numbers at cavity $A$ and cavity $B$ are $\text{Kn}_A=0.001$ and $\text{Kn}_B=10$, and the corresponding pressures are $P_A=48.78\text{Pa}$ and  $P_B=0.04878\text{Pa}$, respectively.
At time $t=0$, the diaphragm is removed suddenly, and then the gas starts to expand from the left cavity to the right one.
We are interested in the dynamic behavior of the gas during the expansion process.

\begin{figure}[htbp]
\centering
\includegraphics[width=0.85\textwidth]{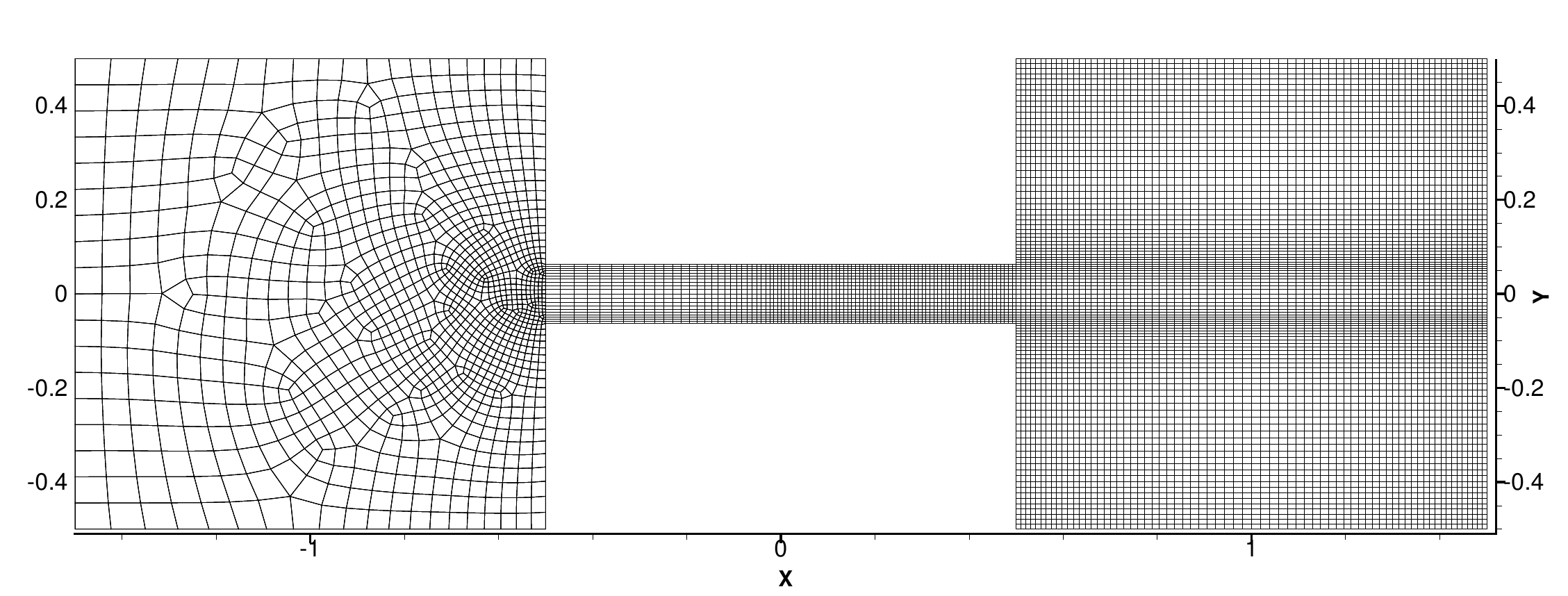}
\caption{Mesh for the gas expansion case}
\label{fig:mesh_expansion}
\end{figure}

The mesh for this case is shown in Fig.~\ref{fig:mesh_expansion}.
As significant flow variations can take place in cavity $B$, the mesh is much finer there. While in cavity $A$, the flow changes slowly and the mesh is relatively coarser.
Note that like the continuum cavity flow, the cell size in cavity $A$ is much larger than the  mean free path there. The correctness of using such a coarser mesh in cavity $A$ is granted by the AP property of the DUGKS.
To account for the highly non-equilibrium effect in cavity $B$ at the early stage,
we use a $101\times101$ mesh distributed uniformly in the range of $[-7\sqrt{2RT_\text{w}},7\sqrt{2RT_\text{w}}]\times[-7\sqrt{2RT_\text{w}},7\sqrt{2RT_\text{w}}]$ for the velocity space discretization, and the Newton-Cotes quadrature rule is used for the numerical integration.
Note that a wider bound is used for the discrete velocity than that used in the cavity flow simulations to account for the  supersonic flow behavior in the channel and cavity $B$ in the early stage of expansion. In our simulations, the CFL number is set to be 0.8.

\begin{figure}[htbp]
\centering
\includegraphics[width=0.85\textwidth]{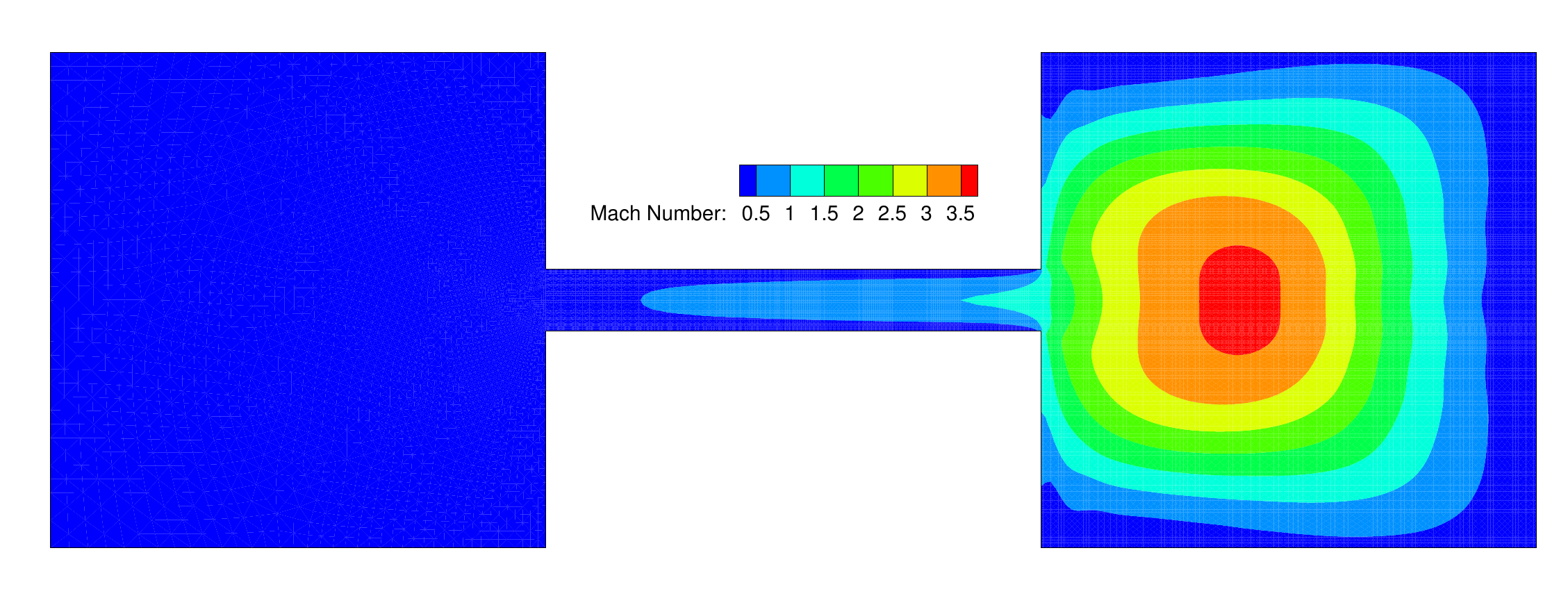}
\caption{Mach number contours for the gas expansion problem at time $t/t_c=1$}
\label{fig:ex_ma_t1}
\end{figure}

\begin{figure}[htbp]
\centering
\includegraphics[width=0.85\textwidth]{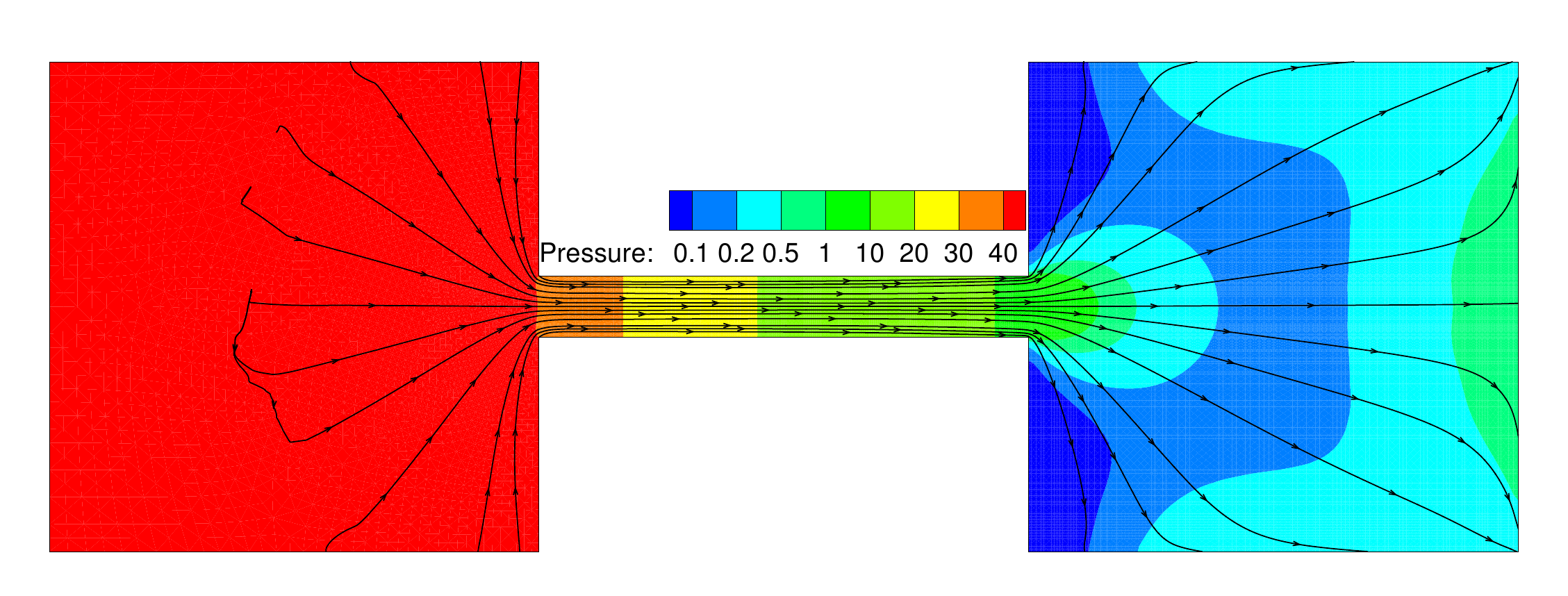}
\caption{Pressure contours and streamlines for the gas expansion problem at time $t/t_c=1$}
\label{fig:ex_P_t1}
\end{figure}

\begin{figure}[htbp]
\centering
\includegraphics[width=0.85\textwidth]{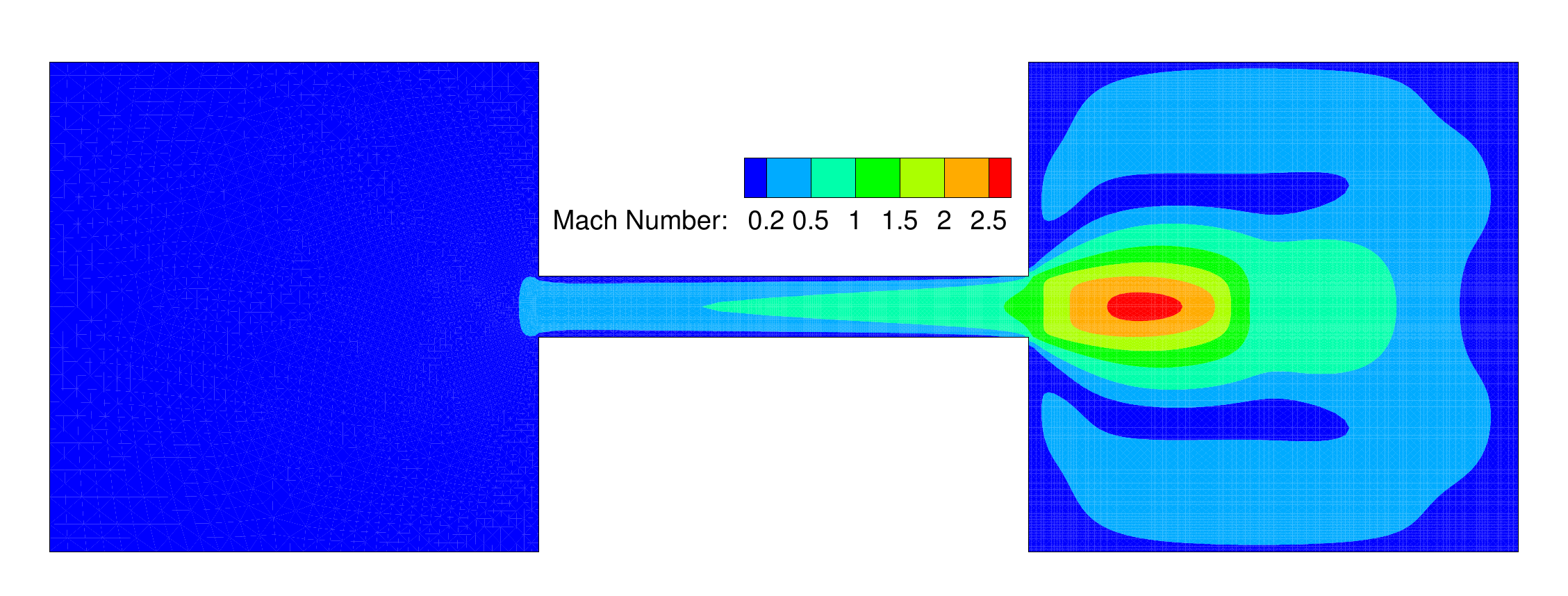}
\caption{Mach number contours for the gas expansion problem at time $t/t_c=4$}
\label{fig:ex_ma_t4}
\end{figure}

\begin{figure}[htbp]
\centering
\includegraphics[width=0.85\textwidth]{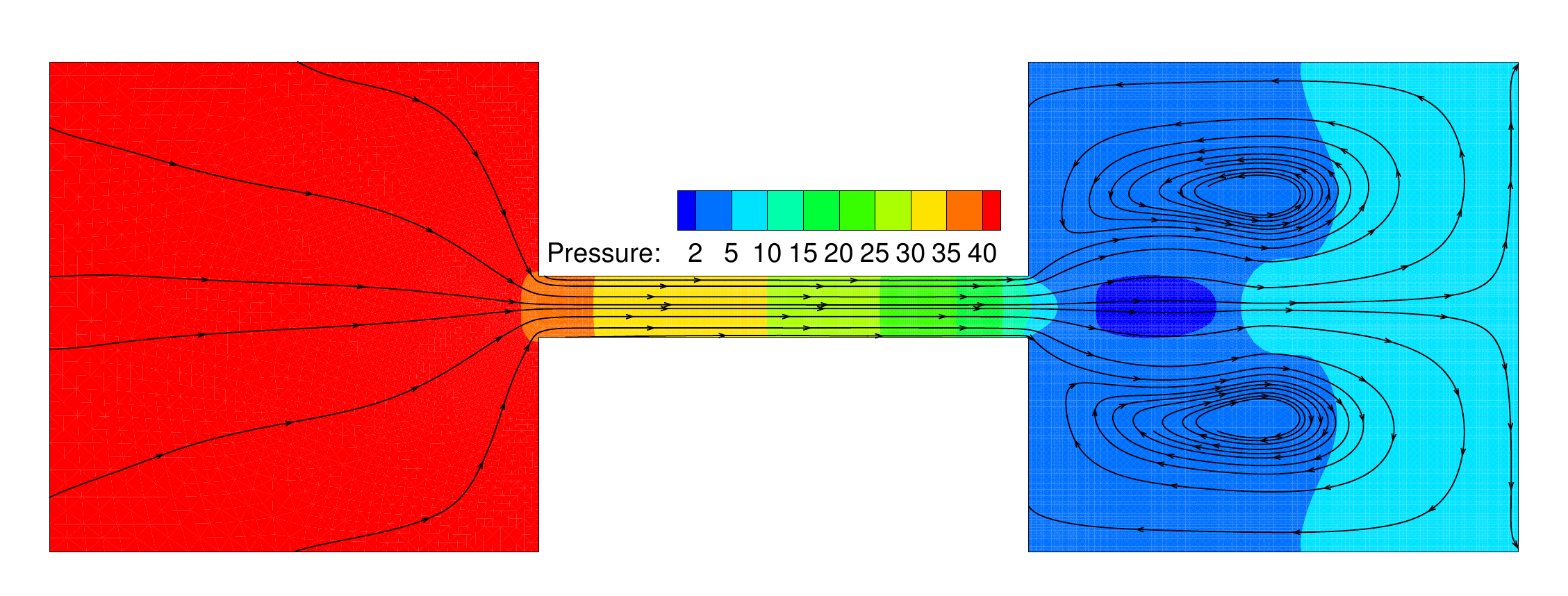}
\caption{Pressure contours and streamlines for the gas expansion problem at time $t/t_c=4$}
\label{fig:ex_P_t4}
\end{figure}

We first define a characteristic time of the system as $t_c=L/\sqrt{2RT_\text{w}}$, and the flow fields at  different times are measured. The local Mach number, pressure, and streamlines at times $t/t_c=1$ and $4$ are presented in Figures~\ref{fig:ex_ma_t1}-\ref{fig:ex_P_t4}.
From Fig.~\ref{fig:ex_ma_t1} and Fig.~\ref{fig:ex_P_t1} we can see that  
the shock wave just reaches the center of cavity $B$ at time $t/t_c=1$. At this time the gas is still very rarefied there, such that the viscous effect can be neglected
without vortex formation.
As the gas flows into cavity $B$ gradually, the pressure in cavity $B$ grows up with time, but the pressure ratio between the two cavities is still high enough to form a supersonic nozzle flow in the channel, and the initial shock wave disappears and two symmetry vortexes appear in cavity $B$ at a later time.

\begin{figure}[htbp]
\centering
  \subfloat[]{\includegraphics[width=0.48\textwidth]{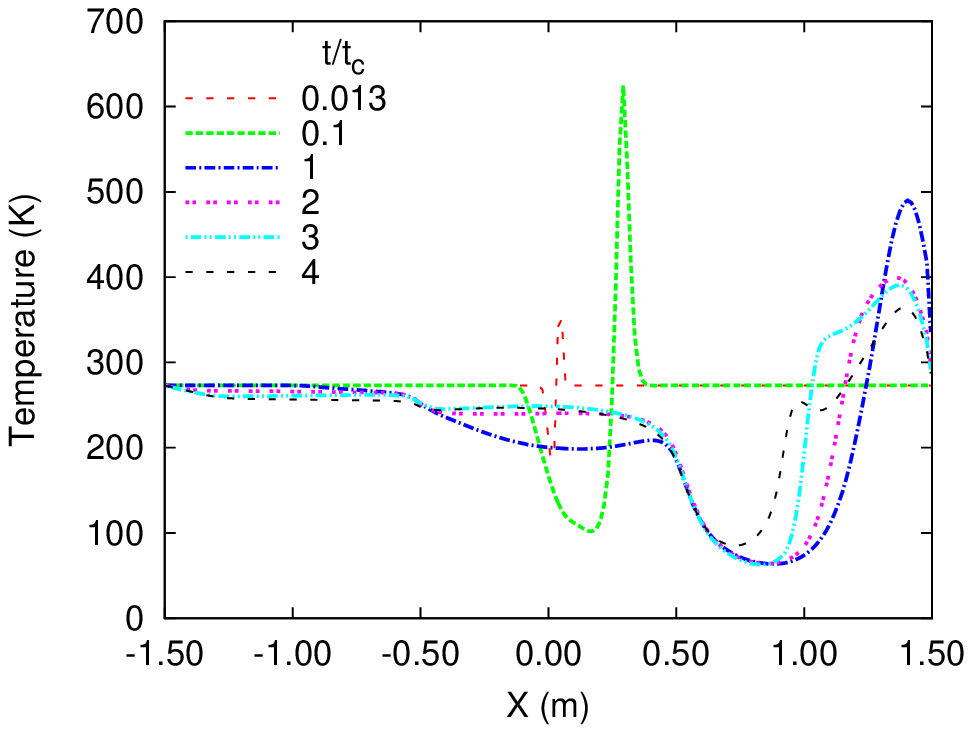}}~
  \subfloat[]{\includegraphics[width=0.48\textwidth]{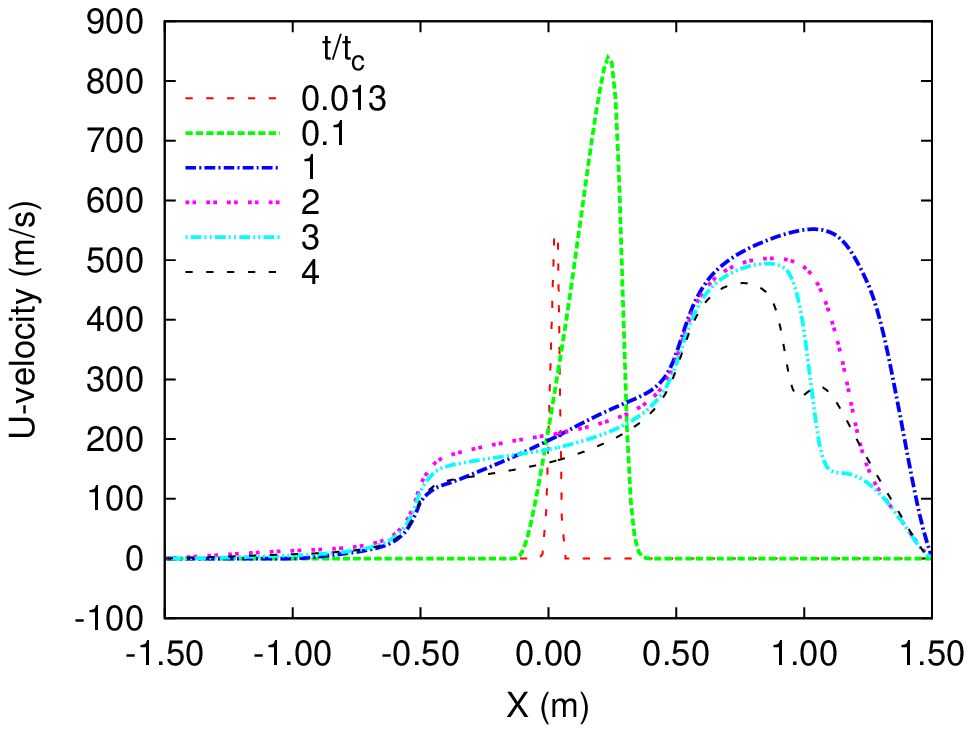}}\\
  \subfloat[]{\includegraphics[width=0.48\textwidth]{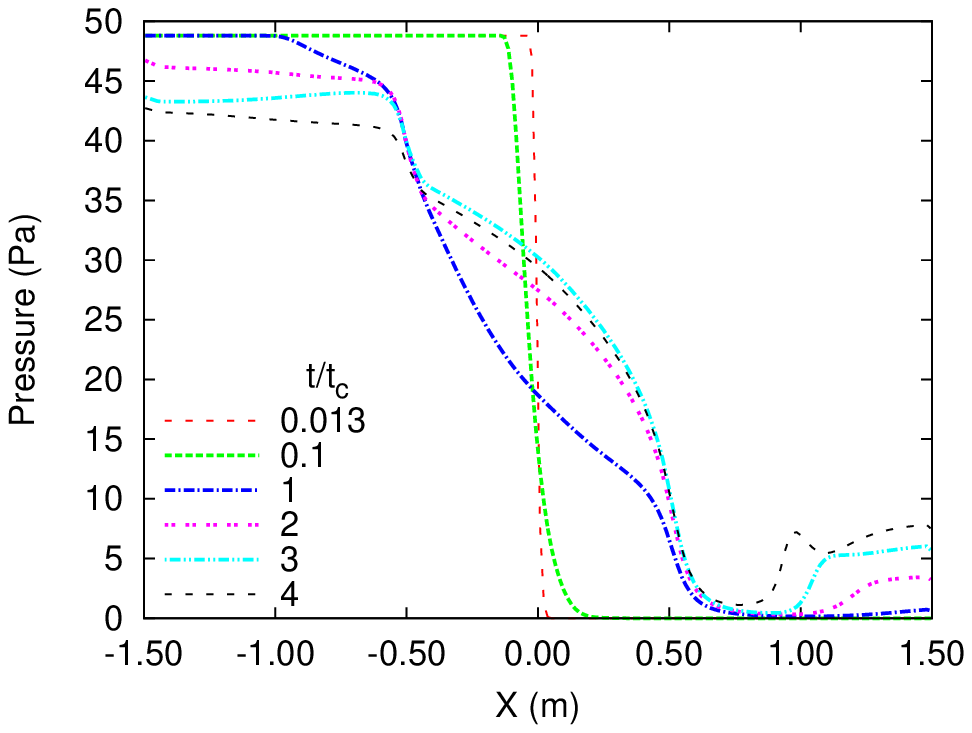}}~
  \subfloat[]{\includegraphics[width=0.48\textwidth]{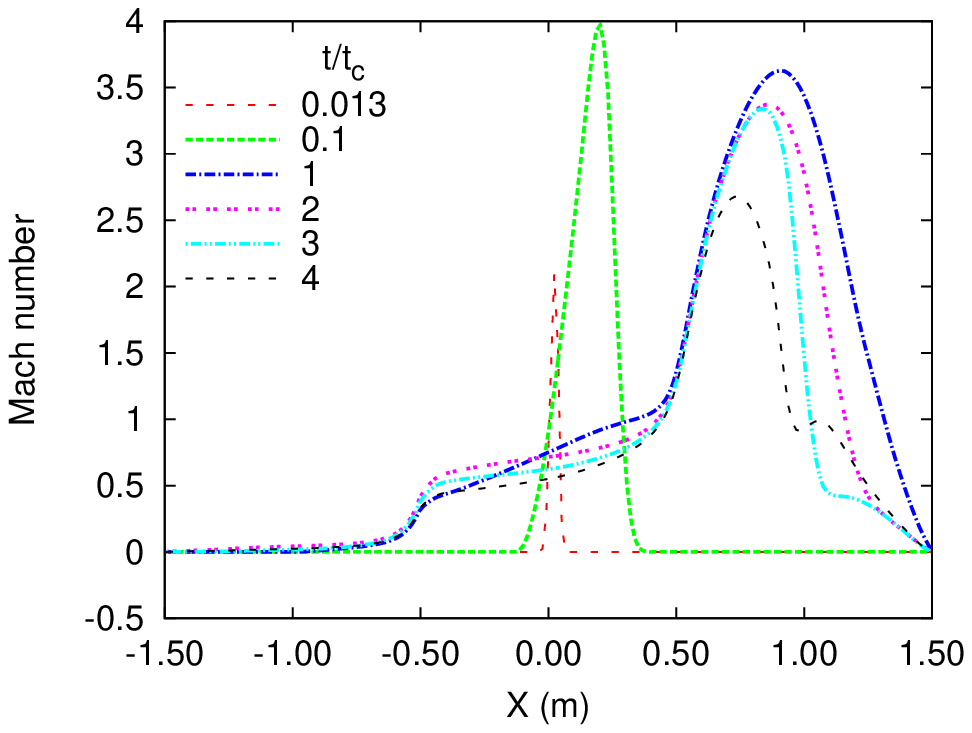}}
\caption{
Temperature (a), horizontal velocity (b), pressure (c) and Mach number (d) 
along the horizontal center line across the cavities and the channel at different times for the gas expansion problem. 
} \label{fig:ex_time}
\end{figure}

To get a detail information of the evolution process of the expansion,
we show the temperature, U-velocity and pressure profiles along the horizontal center line of the cavities and the channel at 
times $t/t_c=0.013, 0.1, 1, 2, 3$ and $4$ in Fig.~\ref{fig:ex_time}.
It can be seen that at the early stage ($t/t_c=0.013$ and $0.1$),
the shock wave propagates in the channel, and the flow variables change sharply across the shock. With time advancing, the pressure difference between the two cavities decreases, and the shock becomes weaker, and the flow rate decreases gradually.

\begin{figure}[htbp]
\centering
\subfloat[]{\includegraphics[width=0.48\textwidth]{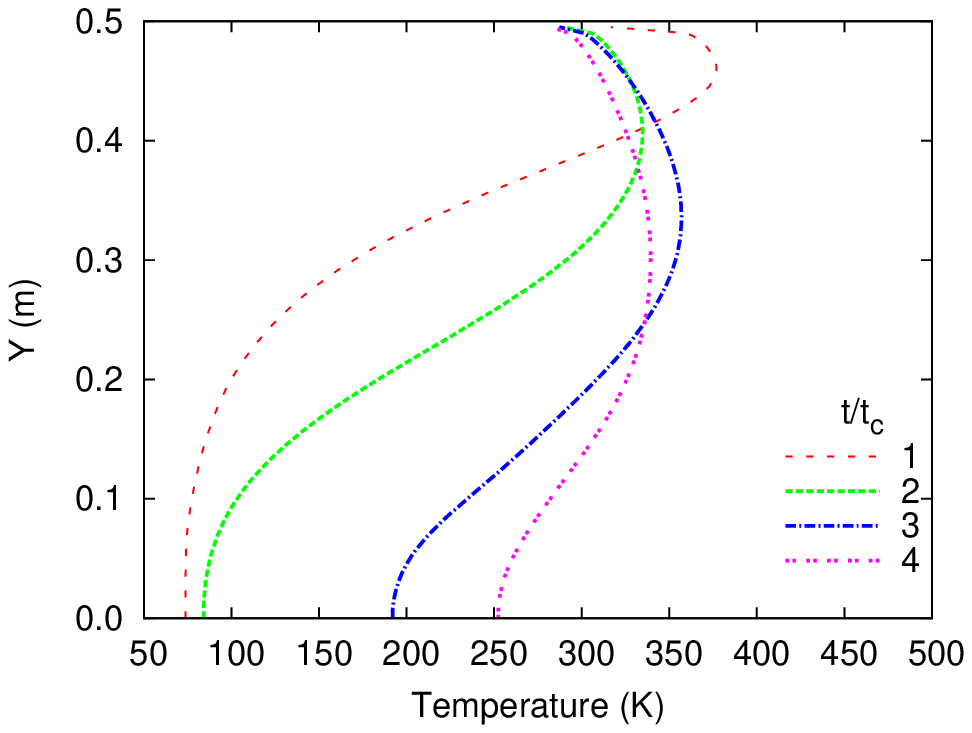}}~
\subfloat[]{\includegraphics[width=0.48\textwidth]{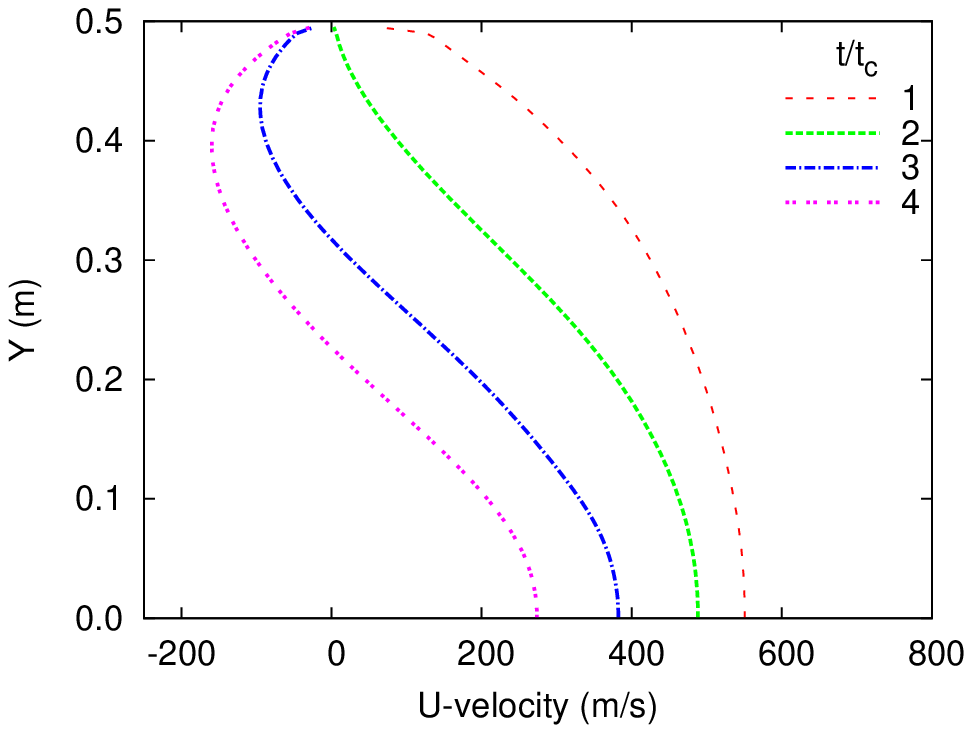}}\\
\subfloat[]{\includegraphics[width=0.48\textwidth]{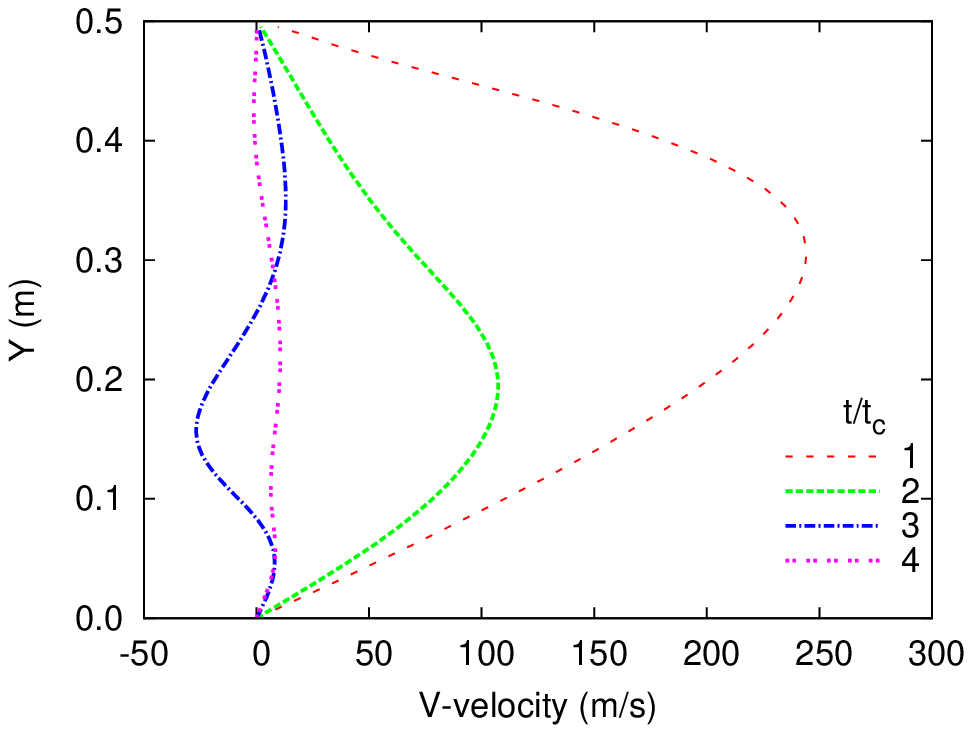}}~
\subfloat[]{\includegraphics[width=0.48\textwidth]{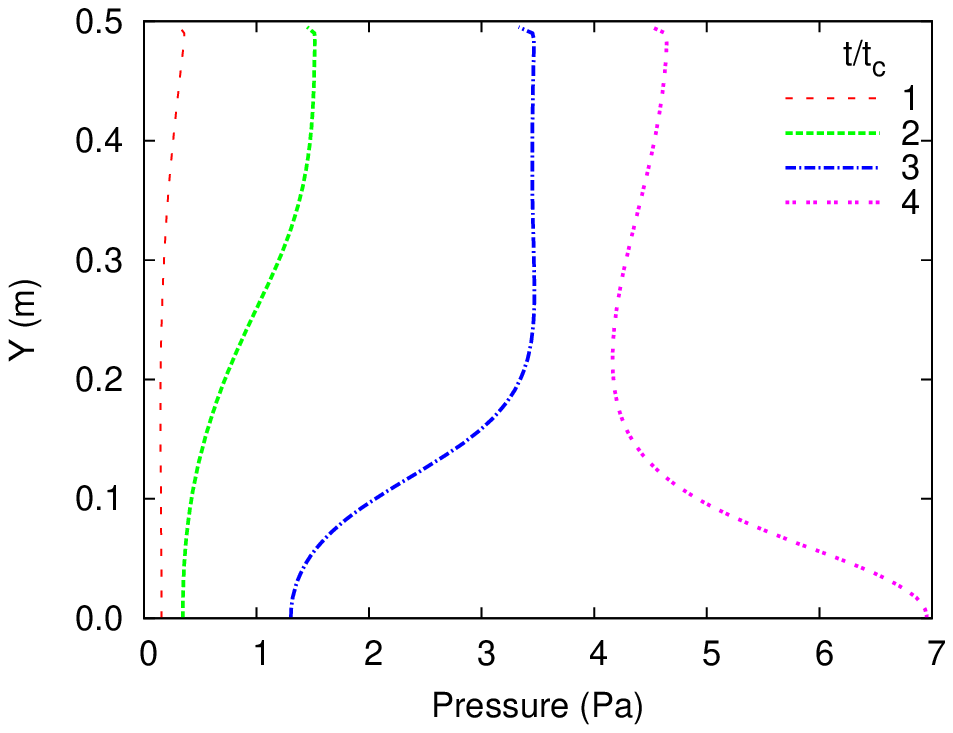}}
\caption{
 Temperature (a), horizontal velocity (b), vertical velocity (c) and pressure (d) along the vertical center line (upper half) of the cavity $B$ at different times for the gas expansion problem. 
} \label{fig:ex_xB_time}
\end{figure}

\begin{figure}[htbp]
\centering
\subfloat[]{\includegraphics[width=0.48\textwidth]{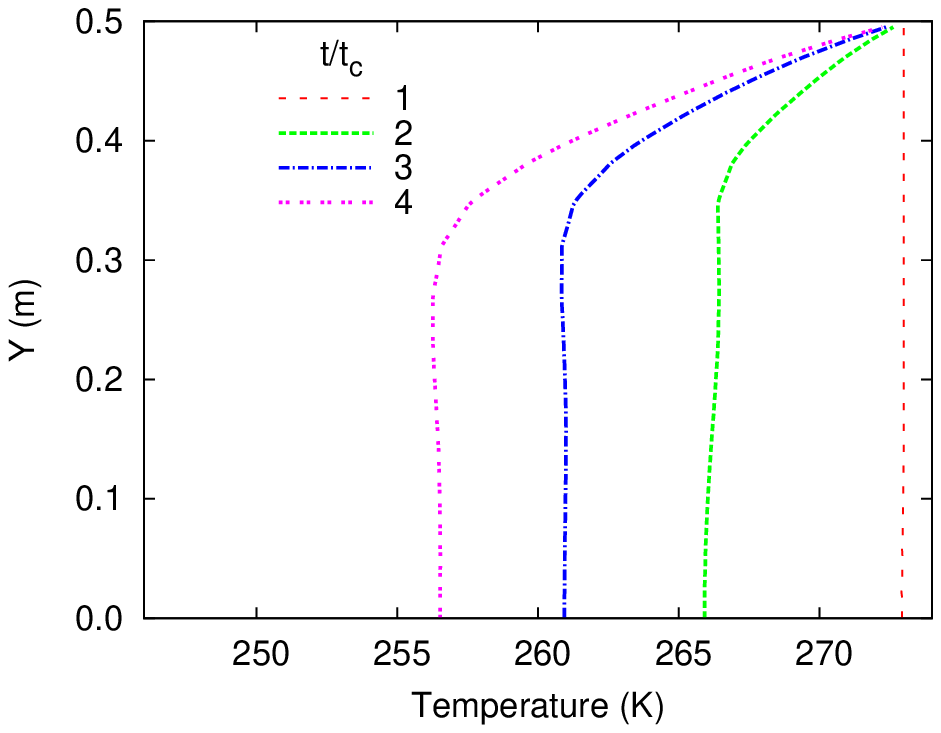}}~
\subfloat[]{\includegraphics[width=0.48\textwidth]{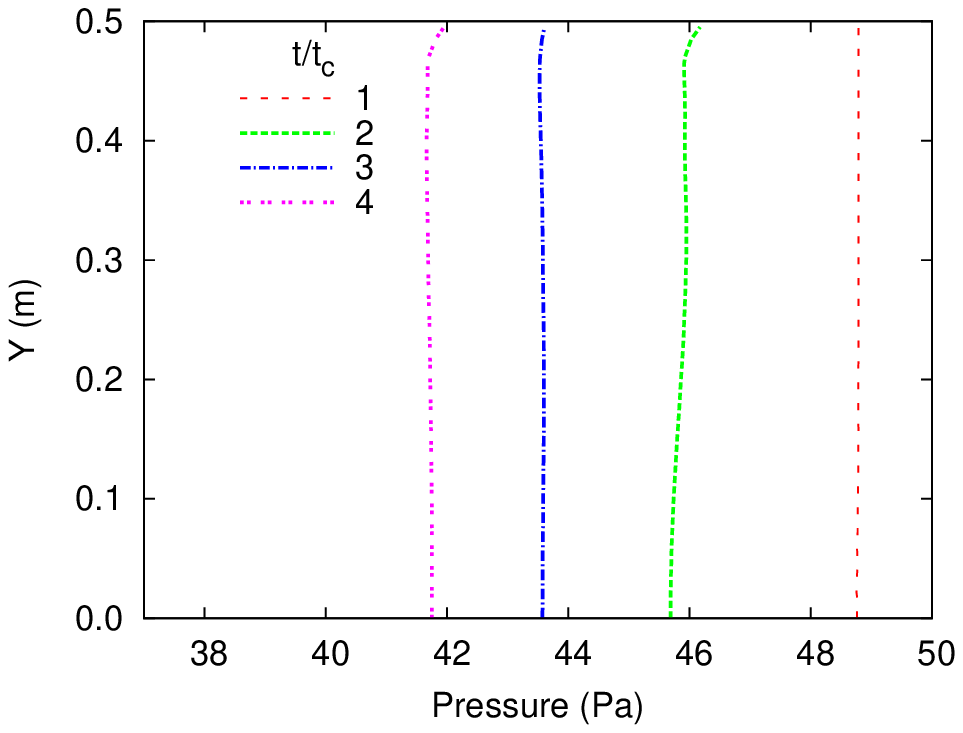}}
\caption{
Temperature (a) and pressure (b) along the vertical center line (upper half) of the cavity $A$ at different times for the gas expansion problem.
} \label{fig:ex_xA_time}
\end{figure}

To quantify the results, the temperature, velocity and pressure profiles along the vertical center lines of the two cavities at different times are presented in Fig.~\ref{fig:ex_xB_time} and Fig.~\ref{fig:ex_xA_time}, respectively. Here only the results at the upper half ( $0 < y < L/2$ ) of the domain  are shown owing to the symmetry of the flow.
From Fig.~\ref{fig:ex_xB_time}(b), we can see that a counterclockwise vortex develops in the upper half of cavity $B$, which enhances heat convection in the gas. Consequently, the temperature field becomes uniform gradually, as indicated in Fig.~\ref{fig:ex_xB_time}(a).

On the other hand, the flow in cavity $A$ changes only slightly. The temperature and pressure are almost uniform at each time. With the decreasing of pressure in the cavity, the temperature reduces as the internal energy being converted to the kinetic energy.

\subsection{Supersonic flow passing through a circular cylinder}
To further demonstrate the performance of the DUGKS on unstructured meshes for high speed non-equilibrium external flows, we simulate the rarefied gas flows passing through a circular cylinder. It is noted that this problem was also studied by Huang \emph{et~al.} \cite{huangjc12} using the UGKS method.
We here adopt the same configuration and parameters as in their simulations.
The free-stream Mach number is $\text{Ma}_\infty=5$, and the radius of the cylinder which is $r=0.01$m. Two Knudsen numbers are considered ($\text{Kn}_\infty=\lambda_\infty / r = 0.1$ and $1$).
The free-stream gas temperature is $T_\infty=273\text{K}$ and is used as the referenced temperature.
The surface of the cylinder maintains a constant temperature at $T_\text{w}=273\text{K}$, and full diffusive boundary condition is assumed.
The outer boundary of the computational domain is a circle with a diameter $D_o = 22r$, and forms a concentric annular with the surface of the cylinder.
The distribution functions coming to the computational domain from the outer boundary are set to the equilibrium state based on the free-stream flow condition.

\begin{figure}[htbp]
\centering
\subfloat[]{\includegraphics[width=0.48\textwidth]{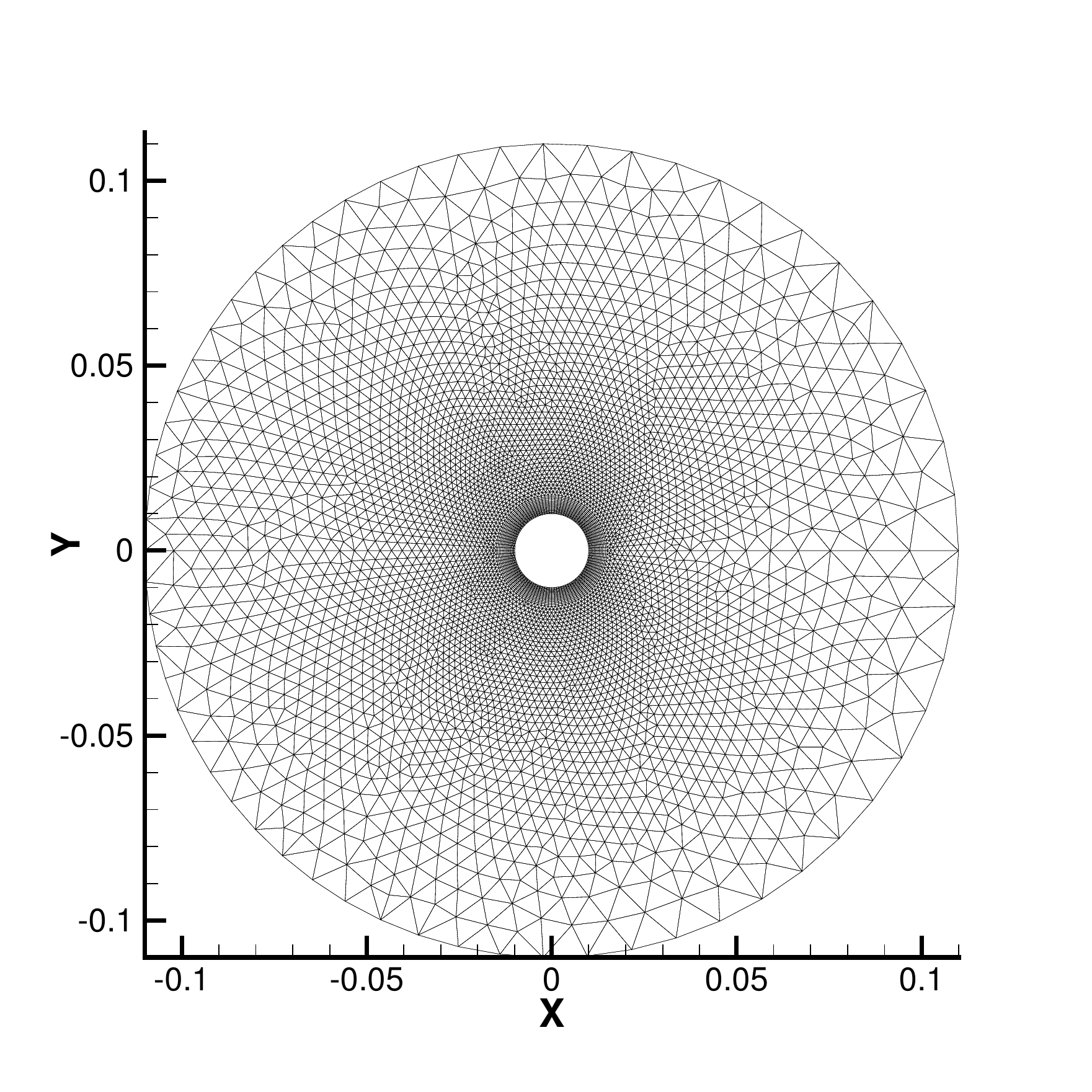}}~
\subfloat[]{\includegraphics[width=0.48\textwidth]{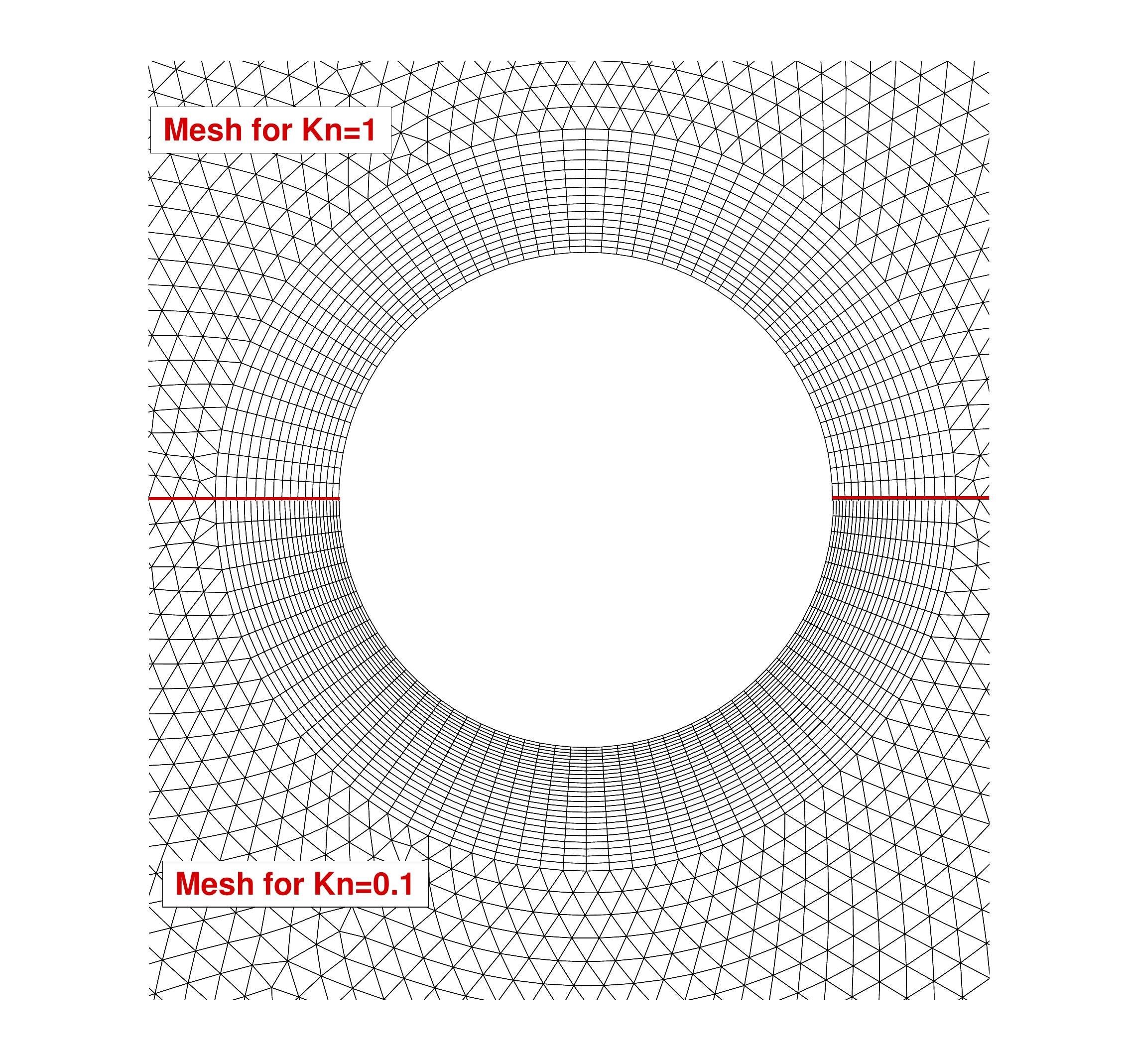}}
\caption{
Meshes for the flow past a cylinder.
(a) Global view of the mesh,
(b) local view of the meshes around the cylinder surface, upper: $\text{Kn}=1$, lower: $\text{Kn}=0.1.$
}\label{fig:cylinder_mesh}
\end{figure}

Hybrid meshes are adopted again for this test case (see Fig.~\ref{fig:cylinder_mesh}). Locally refined quadrilateral cells are used near the cylinder to resolve the boundary layer. We note that the mesh resolution in the normal direction of the cylinder wall should be fine enough near the cylinder to capture the large gradients correctly in the boundary layer.
For the case of $\text{Kn}=0.1$, the mesh spacing around the cylinder wall is finer than that for $\text{Kn}=1$ (see Fig.~\ref{fig:cylinder_mesh}(b)) since the boundary layer become thinner as Kn goes down.
It should be pointed out that the fine resolutions around the cylinder wall are only used to capture the large gradients of the flow field but not to resolve the mean free path scale.
Actually, based on the posterior estimation, the mesh spacing around the stagnation point for the case of $\text{Kn}=1$ is about 2 times that of the mean free path there.

In our computations, the velocity space is discretized into a set of uniform spaced $89 \times 89$ points
in the range of $[-15\sqrt{2RT_\infty}, 15\sqrt{2RT_\infty}] \times [-15\sqrt{2RT_\infty}, 15\sqrt{2RT_\infty}]$, and the
Newton-Cotes quadrature rule is used for the numerical integration.
To validate our simulation results, we use the open source \emph{dsmcFoam} solver \cite{scanlon10} to obtain the DSMC results under the same flow conditions and computational domain.
For the case of $\text{Kn}=1$, the total number of DSMC particles is about 0.58 million and the time step is $1\times 10^{-7}\text{s}$.
For the case of $\text{Kn}=0.1$, the total number of DSMC particles is about 1.11 million and the time step is $4\times 10^{-8}\text{s}$.

\begin{figure}[htbp]
\centering
\subfloat[]{\includegraphics[width=0.48\textwidth]{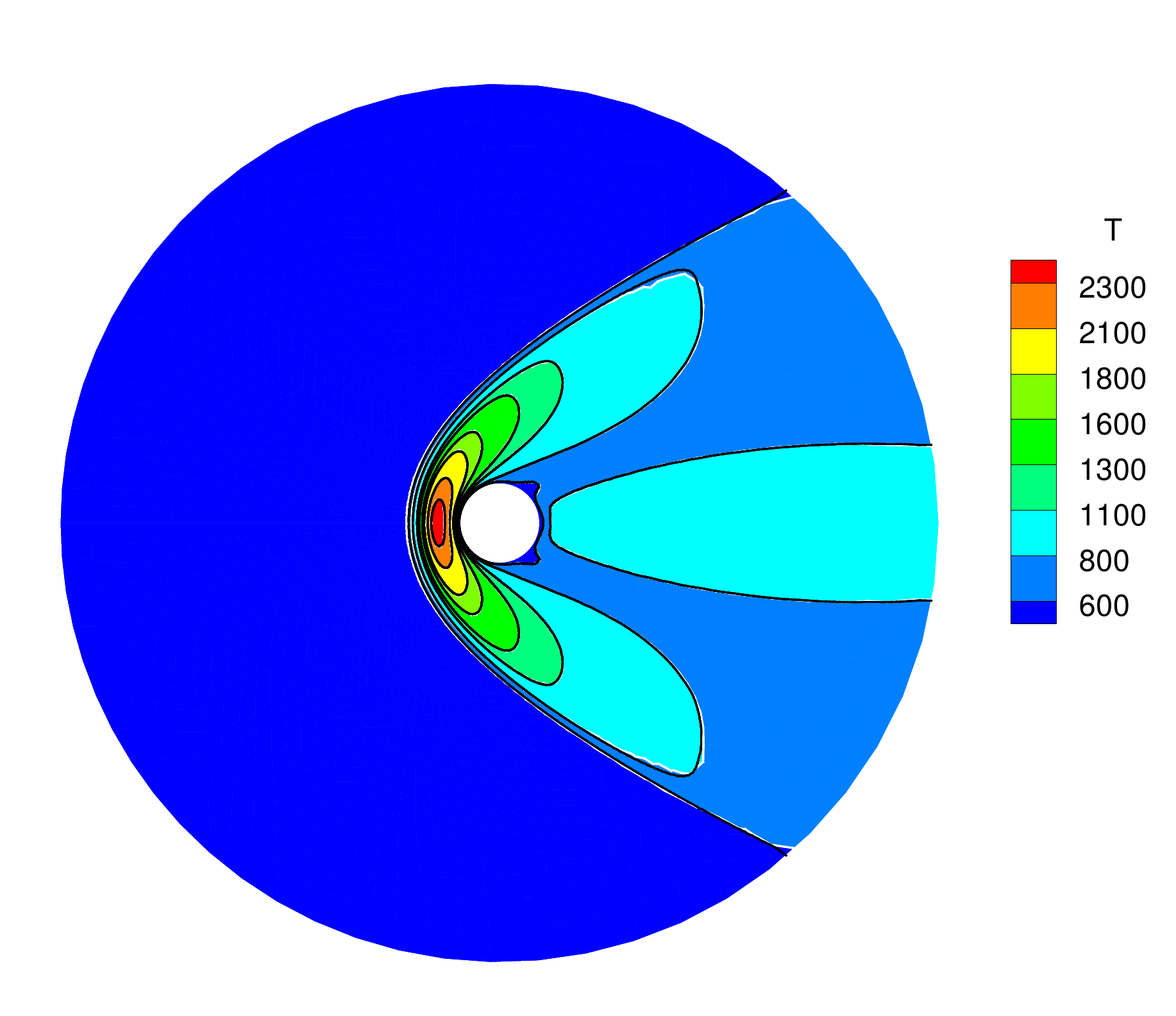}}~
\subfloat[]{\includegraphics[width=0.48\textwidth]{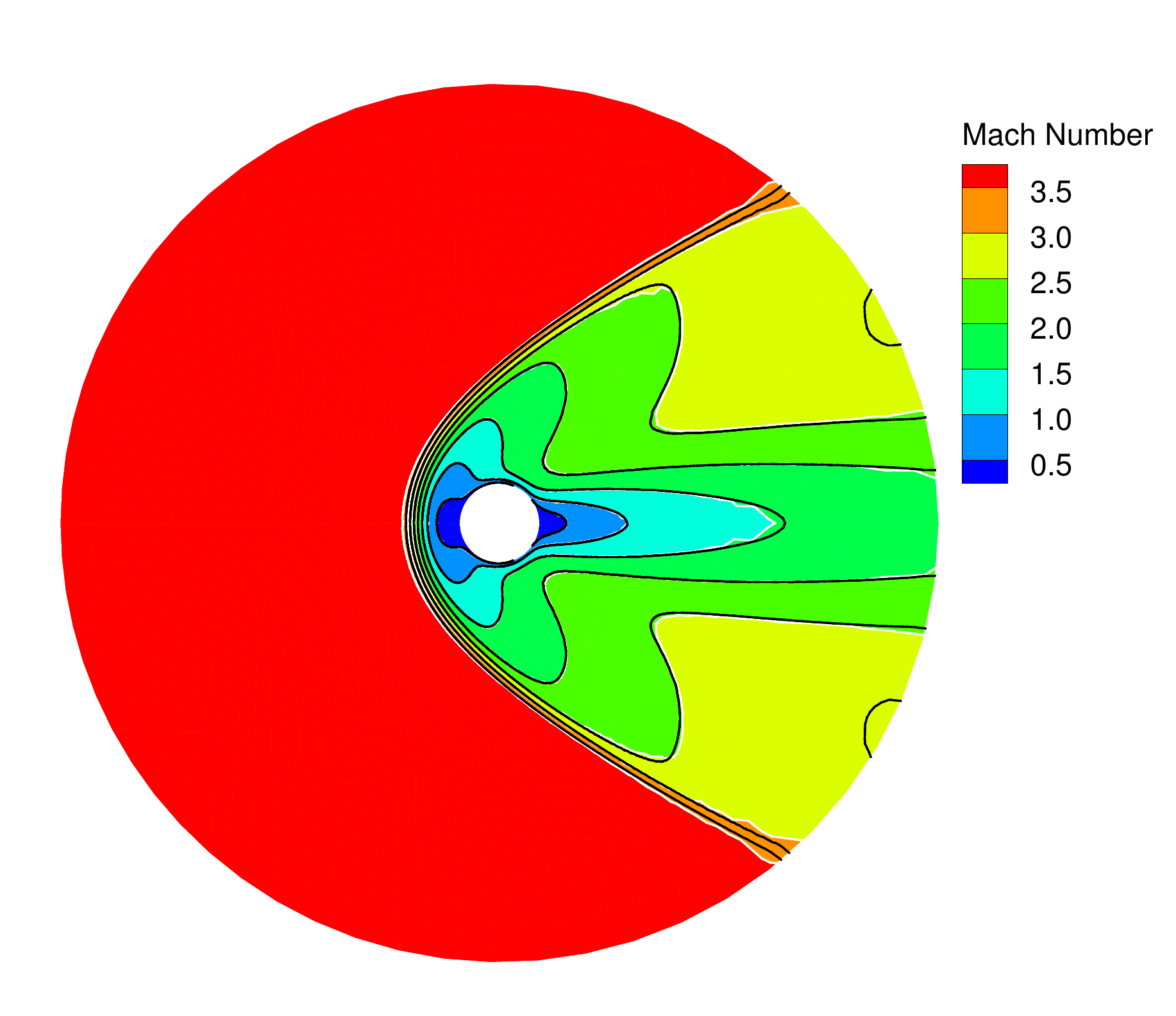}}
\caption{
Temperature (a) and Mach number (b) distribution for the flow past a cylinder at $\text{Kn=0.1}$. Solid white line with colored background: DUGKS, dashed black line: DSMC 
}\label{fig:contour_Kn0d1}
\end{figure}

\begin{figure}[htbp]
\centering
\subfloat[]{\includegraphics[width=0.48\textwidth]{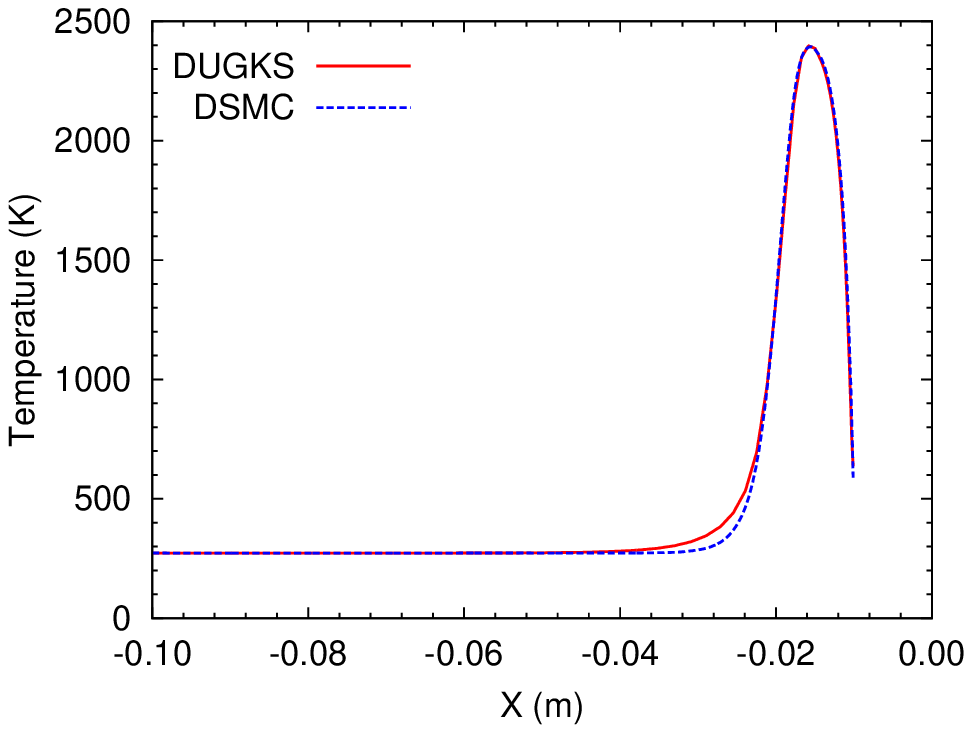}}~
\subfloat[]{\includegraphics[width=0.48\textwidth]{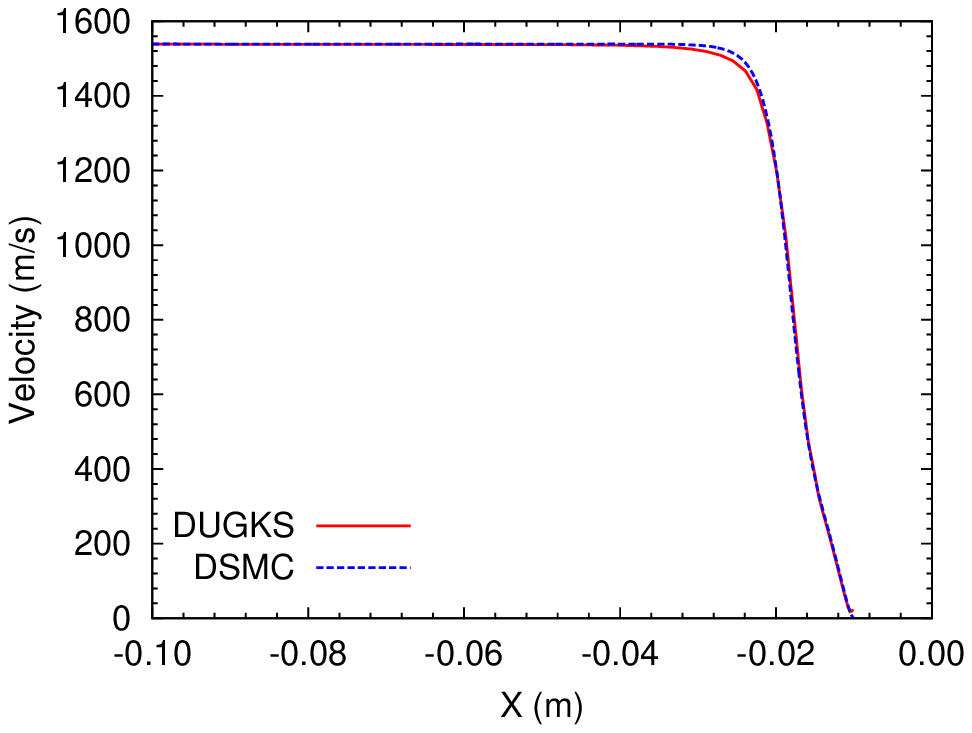}}\\
\subfloat[]{\includegraphics[width=0.48\textwidth]{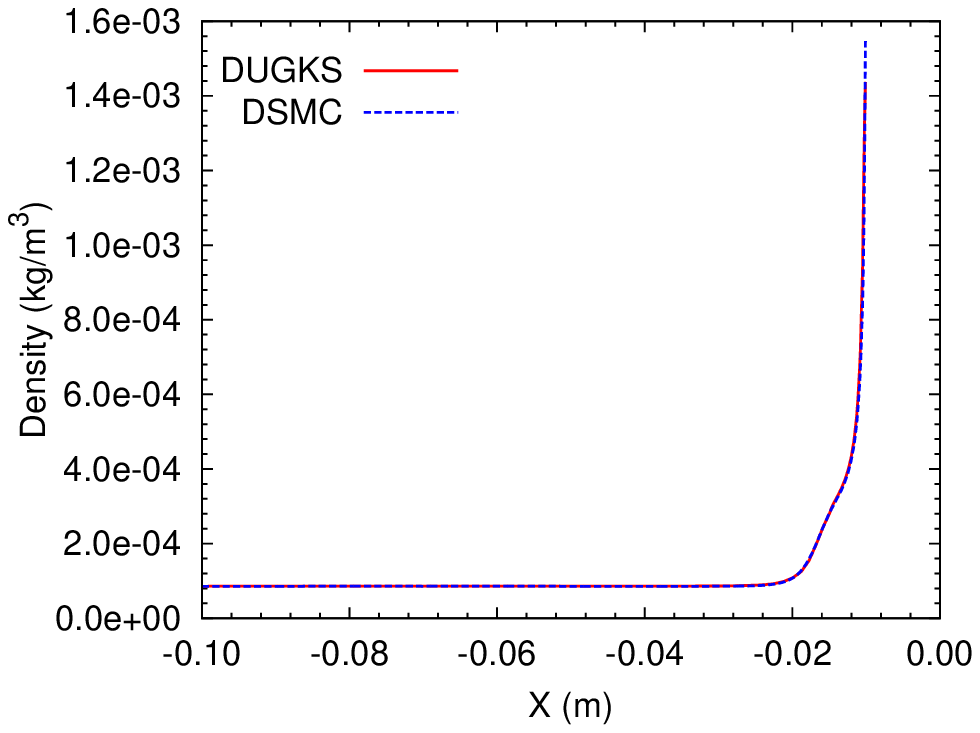}}
\caption{
Temperature (a), velocity (b) and density (c) profiles alone the stagnation line for flow past a cylinder at $\text{Kn=0.1}$. 
} \label{fig:line_Kn0d1}
\end{figure}

\begin{figure}[htbp]
\centering
\subfloat[]{\includegraphics[width=0.48\textwidth]{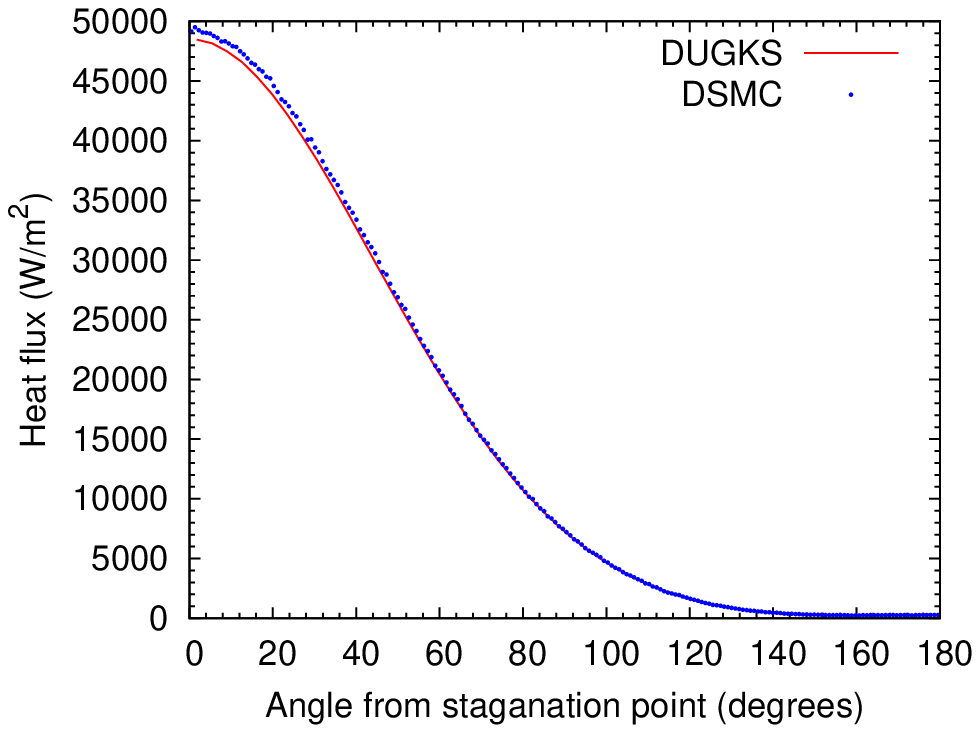}}~
\subfloat[]{\includegraphics[width=0.48\textwidth]{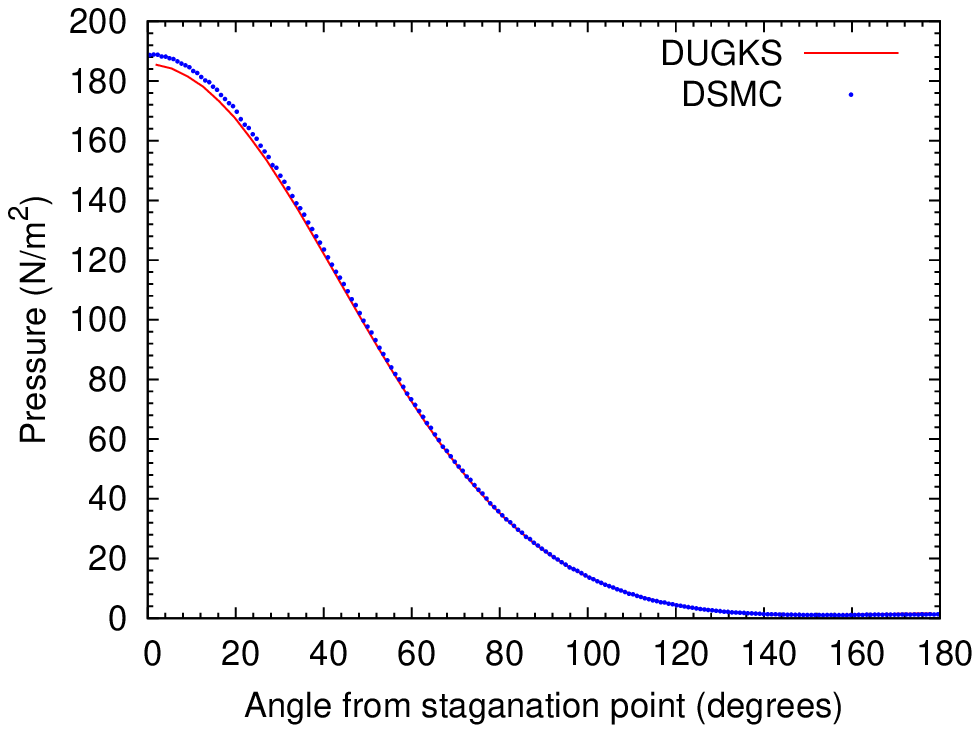}}\\
\subfloat[]{\includegraphics[width=0.48\textwidth]{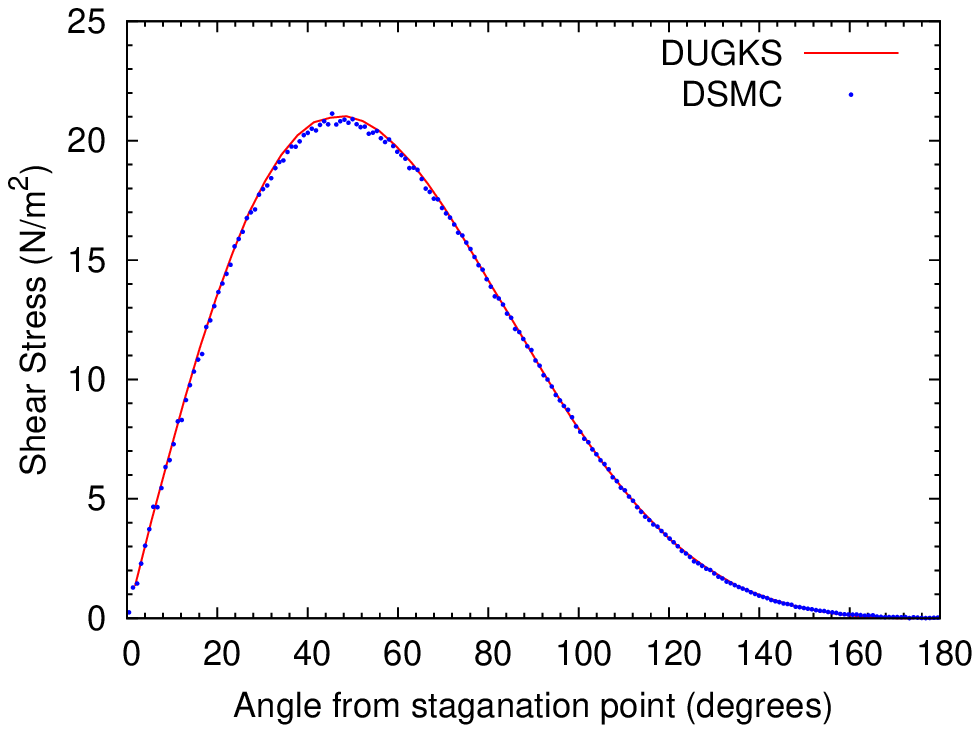}}
\caption{
Heat flux (a), pressure (b) and shear stress (c) alone the surface for the flow past a cylinder at $\text{Kn=0.1}$. } 
\label{fig:wall_Kn0d1}
\end{figure}

The contours of temperature and Mach number for the case of $\text{Kn}=0.1$ are shown in Fig.~\ref{fig:contour_Kn0d1}, also included are the DSMC solutions.
The temperature and U-velocity profile along the stagnation line are shown in Fig.~\ref{fig:line_Kn0d1}.
Clearly we can see that both the temperature and Mach number distributions of the DUGKS results agree with those of the DSMC results perfectly. 
However, there are some discrepancies in the front of the bow shock, which can be seen more clearly in the temperature profile.
This is due to the intrinsic defect of the Shakhov model used in the current DUGKS \cite{bird94, xuk11},
where the collision frequency is independent of particle velocities.
Despite of the small deviations, the temperature agrees well with the DSMC results downstream the shock, and the heat flux,
normal pressure and shear stress distribution along the cylinder's surface predicted by the DUGKS agree with the DSMC results quite well, as shown in Fig.~\ref{fig:wall_Kn0d1}.

\begin{figure}[htbp]
\centering
\subfloat[]{\includegraphics[width=0.48\textwidth]{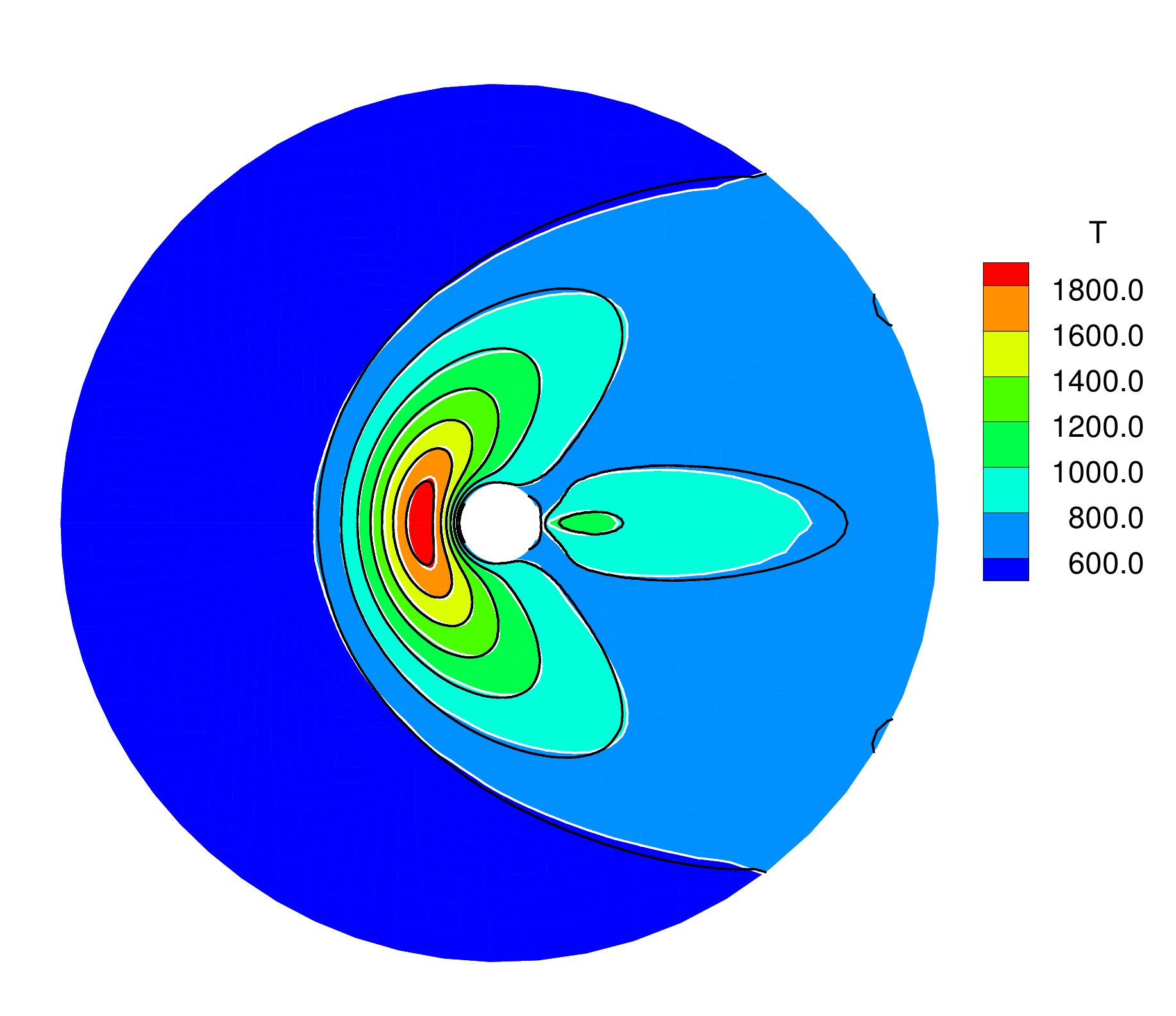}} ~ 
\subfloat[]{\includegraphics[width=0.48\textwidth]{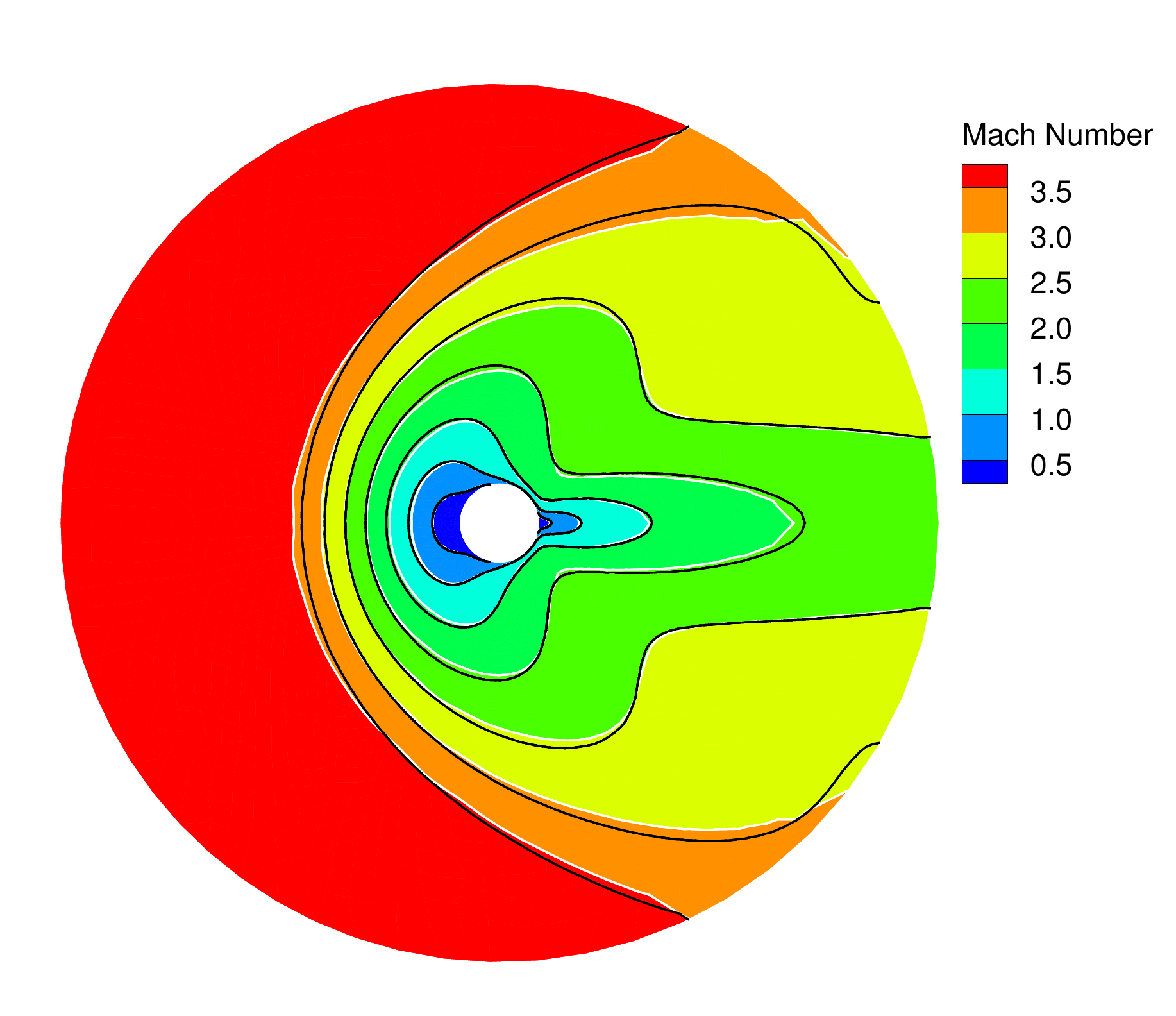}}
\caption{
Temperature and Mach number distributions for the flow past a cylinder at $\text{Kn=1}$. 
Solid white line with colored background: DUGKS, dashed black line: DSMC 
}\label{fig:contour_Kn1}
\end{figure}

\begin{figure}[htbp]
\centering
\subfloat[]{\includegraphics[width=0.48\textwidth]{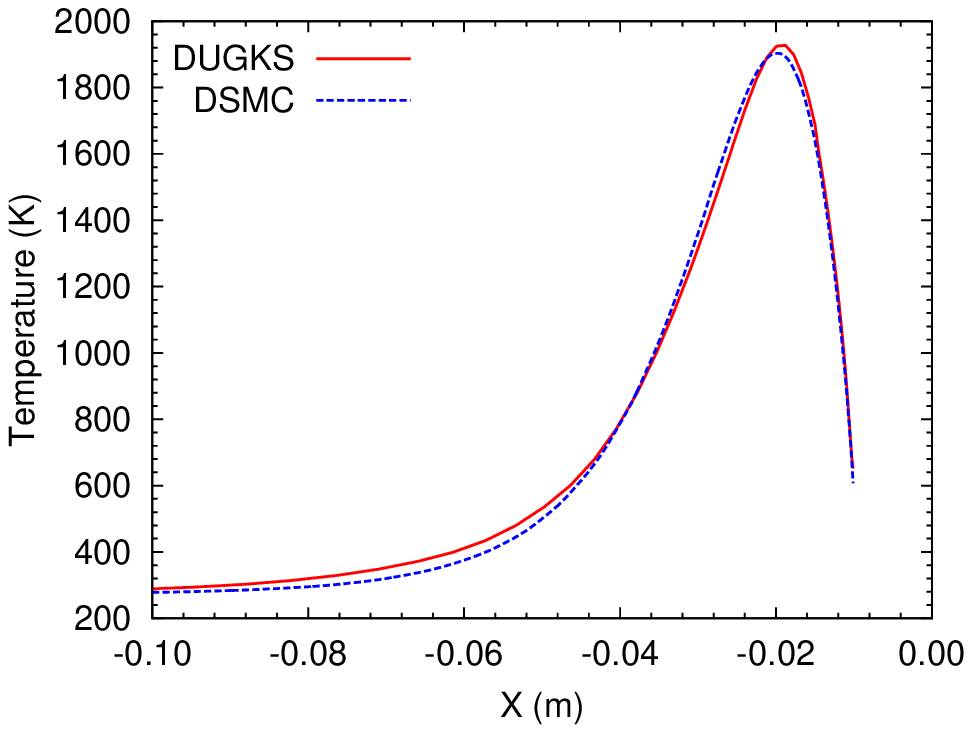}}~
\subfloat[]{\includegraphics[width=0.48\textwidth]{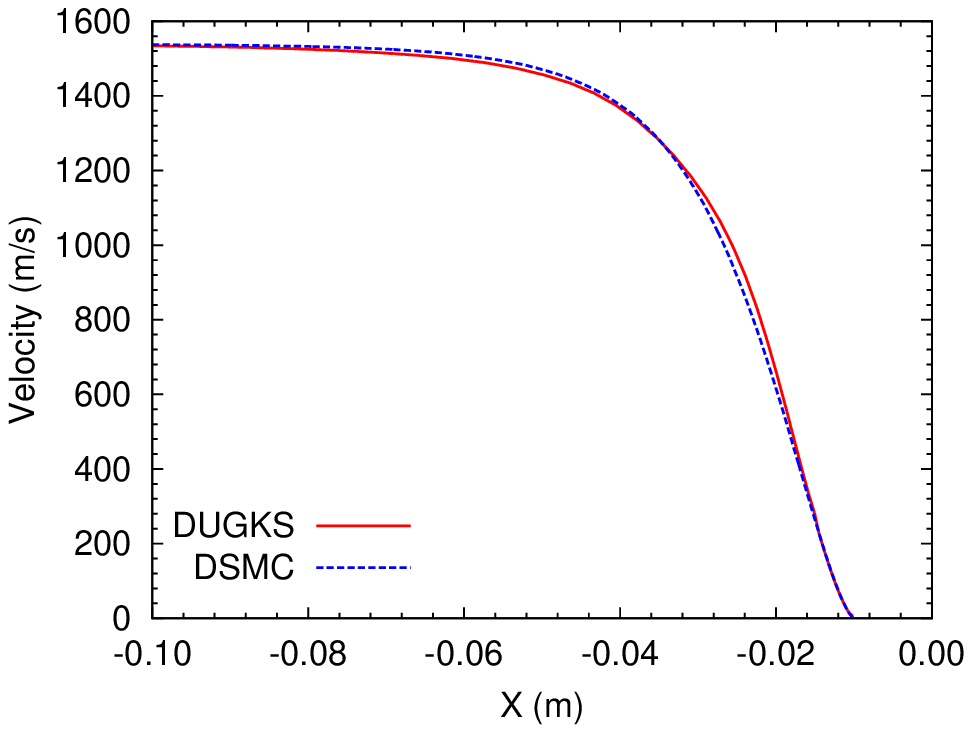}}\\
\subfloat[]{\includegraphics[width=0.48\textwidth]{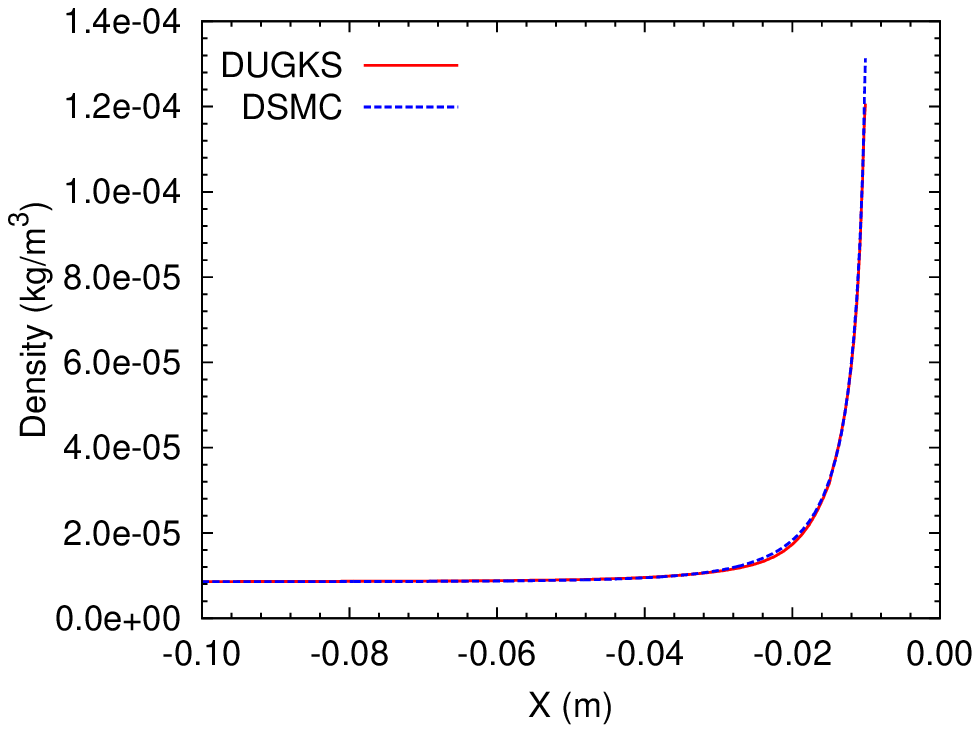}}
\caption{
Temperature (a), velocity (b) and density (c) profiles alone the stagnation line for flow past a cylinder at $\text{Kn=1}$. 
}\label{fig:line_Kn1}
\end{figure}

\begin{figure}[htbp]
\centering
\subfloat[]{\includegraphics[width=0.48\textwidth]{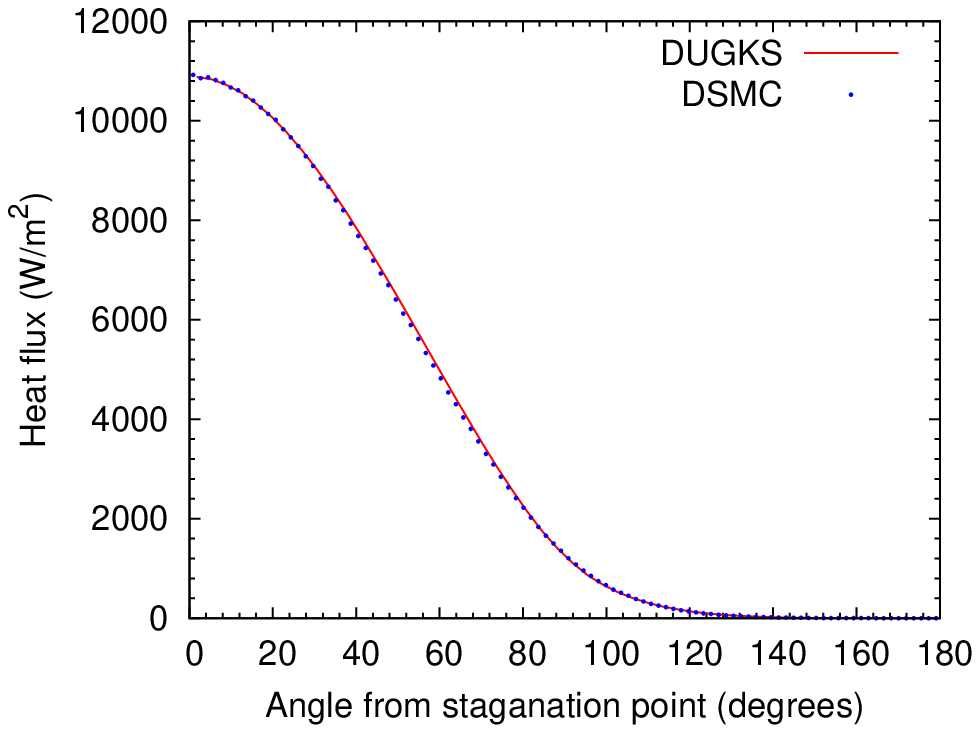}}~
\subfloat[]{\includegraphics[width=0.48\textwidth]{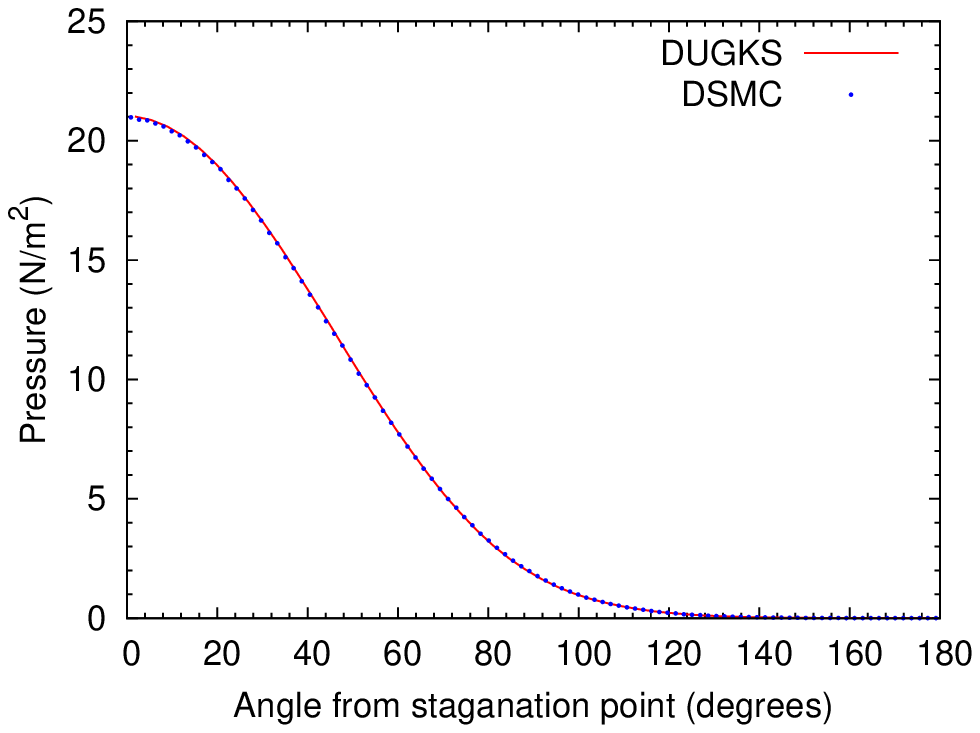}}\\
\subfloat[]{\includegraphics[width=0.48\textwidth]{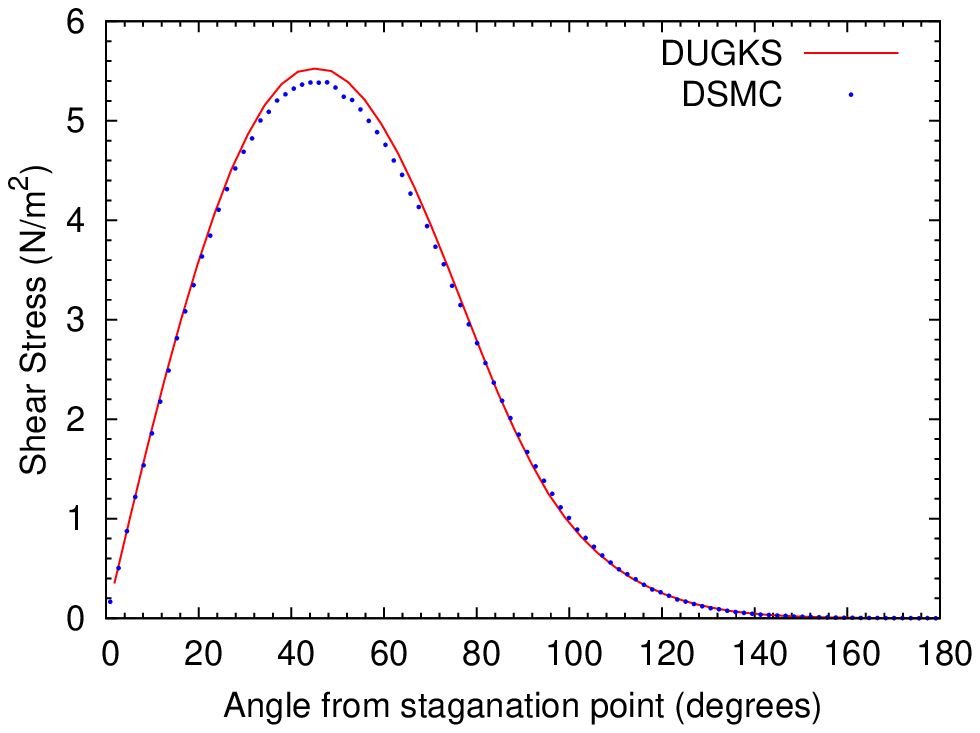}}
\caption{
Heat flux (a), pressure (b) and shear stress (c) alone the surface for the flow past a cylinder at $\text{Kn=1}$. 
}\label{fig:wall_Kn1}
\end{figure}

For the case of $\text{Kn}=1$, 
the temperature and Mach number distributions are presented in Fig.~\ref{fig:contour_Kn1}.
The temperature, U-velocity and density profile along the stagnation line are shown in Fig.~\ref{fig:line_Kn1}. 
These results show that the DUGKS results agree with the DSMC results as well.
The discrepancies in the front of the bow shock are slightly more obvious.
This is because with the increasing of Kn, the non-equilibrium effects get stronger, thus the Shakhov model deviates more from the full Boltzmann collision kernel. However, the heat flux, normal pressure and shear stress along surface of cylinder predicted by DUGKS are still quite satisfactory in comparison with the DSMC results as shown in Fig.~\ref{fig:wall_Kn1}. These results demonstrate that although the Shakhov model has some intrinsic defects, the DUGKS based on it can still give rather satisfactory predictions, particularly the flow behaviors near the body. The DUGKS can be a very useful engineering tool for hypersonic rarefied flow applications, especially in the regime $\text{Kn}<0.1$.

\section{Concluding remarks}
In this paper, the DUGKS based on the Shakhov model developed recently \cite{guozl14} is extended to unstructured meshes.
The key feature of DUGKS is that the discrete characteristic solution of the kinetic equation is used in the modeling of the distribution function at a cell interface.
Due to the coupled treatment of the particle collision and transport processes, the method has the asymptotic preserving (AP) properties for the capturing NS solutions in the continuum flow regime.
Linear reconstruction and gradient limiter are employed in the reconstruction to attain the second-order accuracy.

The performance of the DUGKS on unstructured meshes has been explored by several examples covering different flow regimes from low speed microflows to hypersonic rarefied flows.
In the transitional and slip regimes, good agreements between the results of current scheme and the DSMC solutions are observed;
In the continuum regime, the DUGKS results obtained on a coarse mesh without resolving the mean free path agree with the benchmark solution based on the Navier-Stokes equations very well. 
Thus the AP property of the DUGKS for the Navier-Stokes limit is demonstrated.
The merit of this property is important for a multicale unsteady gas expansion problem that involves both continuum and rarefied regions.
As the mesh size in the continuum region can be much larger than the mean free path scale, 
the overall computational cost for DUGKS can be largely reduced in comparison with the DSMC method and the traditional DVM.
Since the DUGKS is a direct modeling multiscale method \cite{xuk15} method, with the mesh size and time step being a few times of particle mean free path and collision, the physical solutions are not sensitive to individual particle collision. The DUGKS based on kinetic model equation can be faithfully used in real engineering applications.

\section*{Acknowledgments}
We would like to thank Dr. Pubing Yu for helpful discussions.This study is financially supported by National Science Foundation of China (Grant No. 51125024) and Fundamental Research Funds for the Central Universities (Grant No. 2014TS119). K. Xu was supported by Hong Kong Research Grant Council with grants (621011, 620813, 16211014).

\end{document}